\documentclass[a4paper,twoside]{jpconf-mod}
\usepackage{graphicx}
\usepackage{cite}

\def\Lbabar{\mbox{{\LARGE\sl B}\hspace{-0.15em}{\Large\sl A}\hspace{-0.07em}{\LARGE\sl B}\hspace{-0.15em}{\Large\sl A\hspace{-0.02em}R}}}
\def\babar{\mbox{{\small \sl B}\hspace{-0.4em} {\small \sl A}\hspace{-0.03em}{\small \sl B}\hspace{-0.4em} {\small \sl A\hspace{-0.02em}R}}}

\newcommand\T{\rule{0pt}{2.6ex}}       % Top strut
\newcommand\B{\rule[-1.2ex]{0pt}{0pt}} % Bottom strut

\def\ra{\rightarrow} 
\def\CP{$ C \! P$}  
 
\def\gev{$\rm GeV$}
\def\gevc\p{$\rm GeV/c$}
\def\gevc2{$\rm GeV/c^2$}
\def\mev{$\rm MeV$}
\def\mevc{$\rm MeV/c$}
\def\mevc2{$\rm MeV/c^2$}

\def\Cseven{$C_7^{\rm eff}$}
\def\Ceight{$C_8^{\rm eff}$}
\def\Cnine{$C_9^{\rm eff}$}
\def\Cten{$C_{10}^{\rm eff}$}
\def\bsdg{$B \ra  X_{s,d} \gamma$}
\def\bsg{$B \ra  X_s \gamma$}
\def\bdg{$B \ra  X_d \gamma$}
\def\bsdll{$B \ra  X_{s,d} \ell^+ \ell^-$}
\def\bsll{$B \ra  X_s \ell^+ \ell^-$}
\def\bdll{$B \ra  X_d \ell^+ \ell^-$}
\def\bhll{$B^+ \ra X^-  \ell^+ \ell^{\prime +}$}
\def\bdmll{$B^+ \ra D^-  \ell^+ \ell^{\prime +}$}
\def\brholl{$B^+ \ra \rho^-  \ell^+ \ell^{\prime +}$}
\def\bkstll{$B^+ \ra K^{*-}  \ell^+ \ell^{\prime +}$}
\def\bpell{$B \ra \pi /\eta \ell^+ \ell^- $}
\def\bpll{$B \ra \pi  \ell^+ \ell^- $}
\def\bell{$B \ra \eta \ell^+ \ell^- $}
\def\bomom{$B^0 \ra \omega \omega$}
\def\bomfi{$B^0 \ra \omega \phi$}
\def\Csc{$C_S$}
\def\Cps{$C_P$}
\def\imces{$ {\cal I}m(C^{\rm eff}_8 / C^{\rm eff}_7)$}
\def\heff{$H_{\rm eff}$}
\def\Eg{$E^*_\gamma$}
\def\acp{${\cal A}_{C\! P}$}
\def\dacpf{$ \Delta {\cal A}_{C\!P} (B \ra X_s \gamma) =  {\cal A}_{C\!P} (B^+ \ra X^+_s \gamma) - {\cal A}_{C\!P} (B^0 \ra X^0_s \gamma)$}
\def\dacp{$ \Delta {\cal A}_{C\!P} (B \ra X_s \gamma) $}
\def\dchi{$\Delta \chi^2$}
\def\lamse{$\bar \Lambda_{78}$}
\def\qq{$q \bar q$}
\def\BB{$B \bar B$}
\def\mes{$m_{ES}$}
\def\DE{$\Delta E$}
\def\jpsi{$J/\psi$}
\def\psip{$\psi(2S)$}
\def\ee{$e^+ e^-$}
\def\mm{$\mu^+ \mu^-$}
\def\FL{${\cal F}_L$}

\begin{document}
\title{Recent \Lbabar\ Results}

\author{Gerald Eigen, \\ representing the \babar\ collaboration}

\address{Dept. of Physics, University of Bergen, Allegation 55, Bergen, Norway}

\ead{gerald.eigen@ift.uib.no}

\begin{abstract}
We present herein the most recent \babar\ results on direct \CP\ asymmetry measurements in $B \ra  X_s \gamma$, on partial branching fraction and \CP\ asymmetry measurements in $B \ra  X_s \ell^+ \ell^-$, on a search for $B \ra \pi / \eta \ell^+ \ell^-$ decays, on a search for lepton number violation in \bhll\ modes and a study of \bomom\ and \bomfi\  decays.
\end{abstract}
\vskip -0.5cm

\section{Introduction}

The decays $B \ra X_{s,d} \gamma$ and $B \ra  X_{s,d} \ell^+ \ell^-$ (with $\ell^+ \ell^- = e^+ e^- , ~\mu^+  \mu^-$) are flavor-changing neutral-current (FCNC) processes that are forbidden in the Standard Model (SM) at tree level. They occur in higher-order processes and are described by an effective Hamiltonian that factorizes short-distance contributions in terms of scale-dependent Wilson coefficients $C_i (\mu)$~\cite{Wilson} from long-distance contributions expressed by local four-fermion operators ${\cal O}_i$ that define hadronic matrix elements,

\vskip -0.2cm
\begin{equation}
H_{\rm eff} =\frac{4G_F}{\sqrt{2}} \sum_i C_i(\mu) {\cal O}_i.
\label{eq:heff}
\end{equation}
While Wilson coefficients are calculable perturbatively, the calculation of the hadronic matrix
elements requires non-perturbative methods such as the heavy quark expansion~\cite{Isgur, Georgi, Grinstein}.

\begin{figure}[h]
\centering
\vskip -0.3cm
\includegraphics[width=50mm]{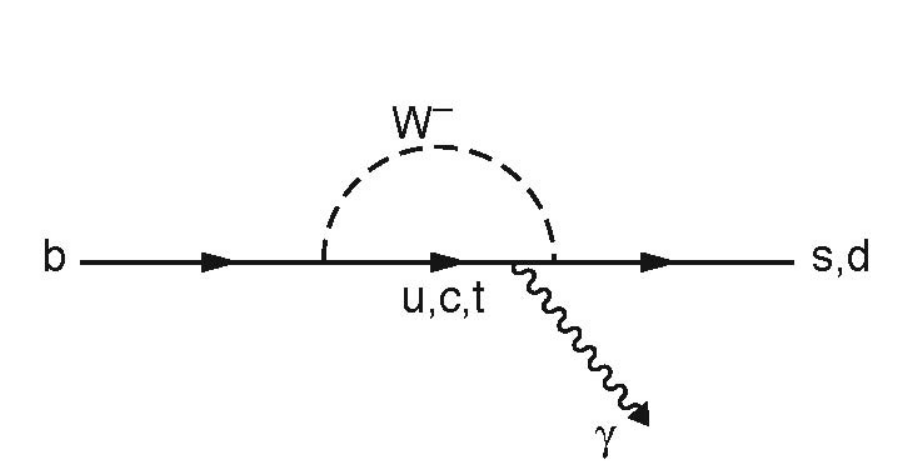}
\includegraphics[width=80mm]{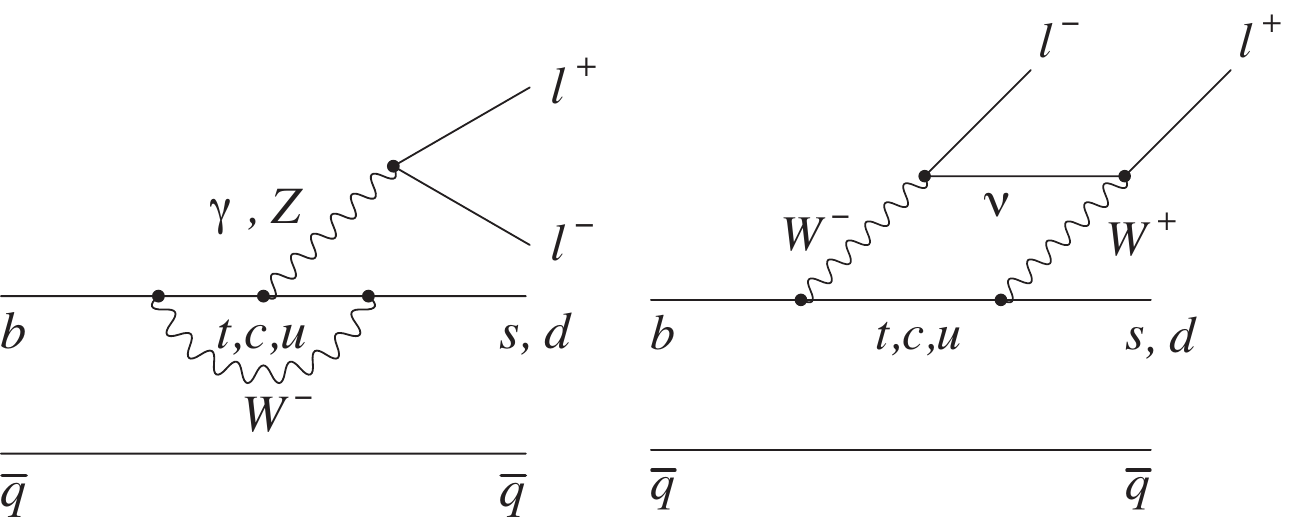}
\vskip -0.2cm
\caption{Lowest-order diagrams for \bsdg\ (left) and \bsdll\  (middle, right).}
\label{fig:penguin}
\end{figure}

\begin{figure}[h]
\centering
\vskip -0.3cm
\includegraphics[width=110mm]{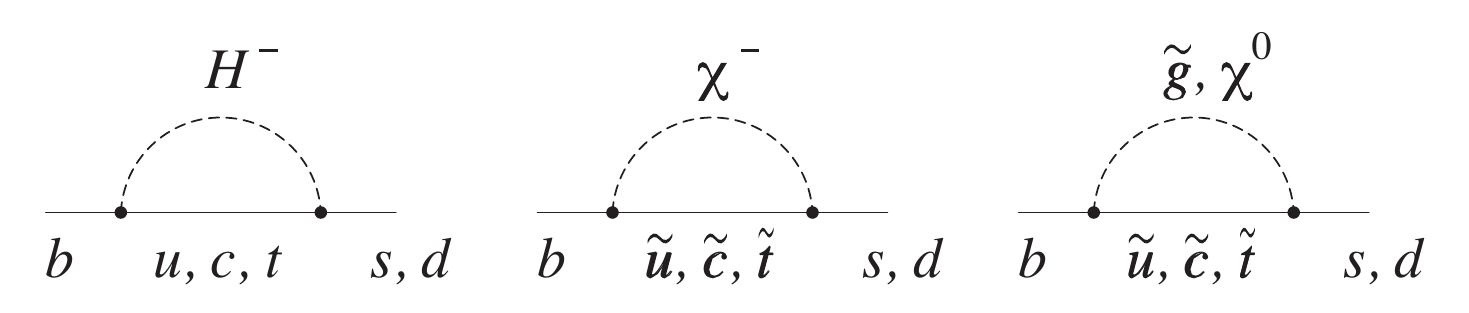}
\includegraphics[width=45mm]{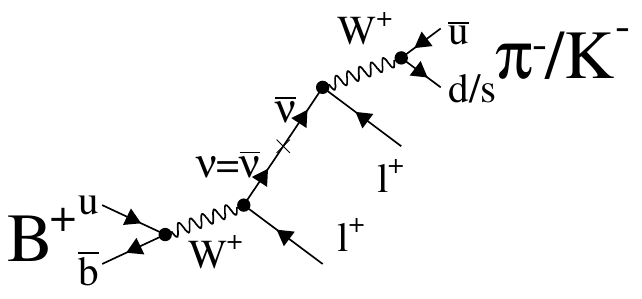}
\vskip -0.2cm
\caption{Examples of new physics processes via a charged Higgs boson (left), charginos (middle
left), neutralinos (middle right) and via Majorana-type neutrino interactions (right).}
\label{fig:np}
\end{figure}

Figure~\ref{fig:penguin} shows the lowest order diagrams for these FCNC decays. In \bsdg, the electromagnetic penguin loop dominates. The short-distance part is expressed by the effective Wilson coefficient \Cseven. Through operator mixing at higher orders, the chromomagnetic
penguin enters whose short distance part is parameterized by \Ceight. In \bsdll  modes, the $Z$ penguin and the $WW$ box diagram contribute in addition whose short-distance parts are parametrized in terms of \Cnine\ and  \Cten, the
vector and axial-vector current  contributions of these diagrams. 
Physics
beyond the SM introduces new loops and box diagrams with new particles (e.g. a charged Higgs
boson or supersymmetric particles) as shown in Fig.~\ref{fig:np} (left, middle). Such contributions modify
the Wilson coefficients and may introduce new diagrams with scalar and pseudoscalar current
interactions and in turn new Wilson coefficients, \Csc\ and \Cps~\cite{Hiller}. To determine \Cseven, \Ceight, \Cnine\ and  \Cten\ precisely, we need to measure many observables in several radiative and rare semileptonic decays, which potentially can probe new physics at a scale of a few TeV.

Lepton-number-violating decays are highly suppressed in the SM and may need new physics
processes. Figure~\ref{fig:np} (right) shows a $W$ annihilation diagram into $\ell^+ \nu_\ell$ in which the neutrino mixes
into an antineutrino producing like-sign leptons that are forbidden in SM interactions.
Such processes require Majorana-type neutrinos that are absent in the SM~\cite{Majorana}.

\begin{figure}[h]
\centering
\vskip -0.5cm
\includegraphics[width=35mm]{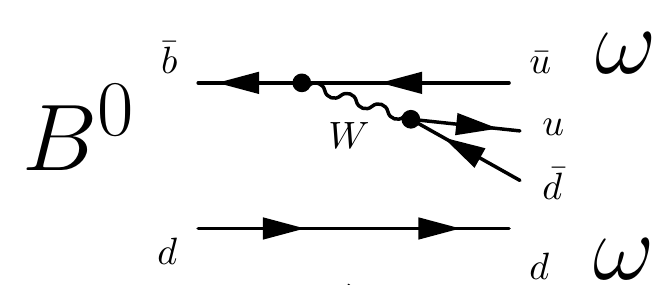}
\includegraphics[width=35mm]{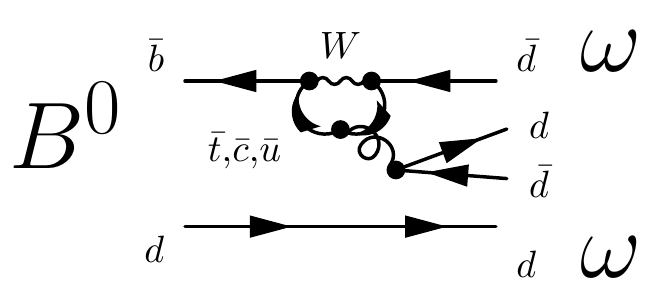}
\includegraphics[width=40mm]{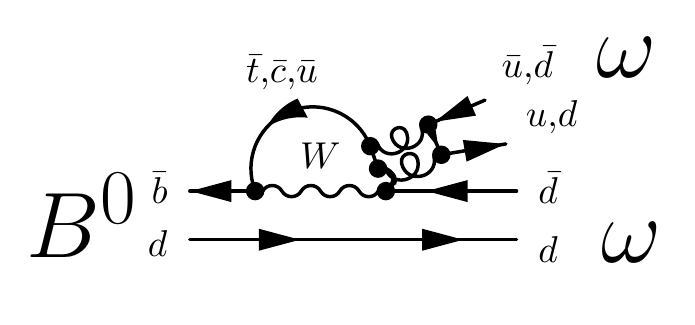}
\includegraphics[width=40mm]{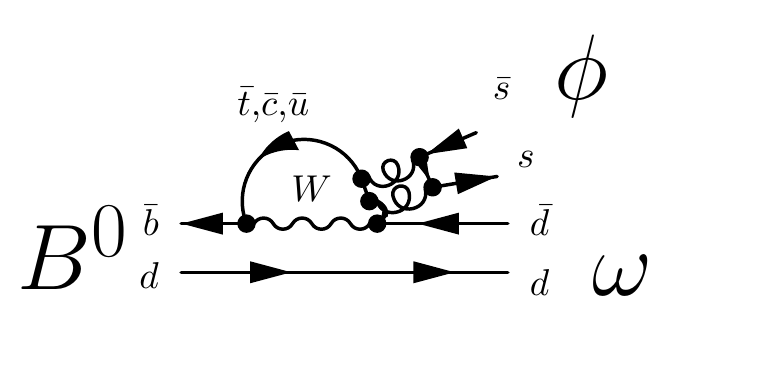}
\vskip -0.2cm
\caption{Lowest-order diagrams for the \bomom\ color-suppressed tree (left), singlet penguin
(middle left) and gluonic penguin (middle right) and the gluonic penguin for \bomfi\ (right).}
\label{fig:omom}
\end{figure}

The decays \bomom\ and \bomfi\ also involve FCNC processes that are mediated by gluonic penguin loops included in \heff\ (see Eqn. (\ref{eq:heff})). Here, the short-distance contributions are
parameterized by the Wilson coefficients $C_3$, $C_4$, $C_5$ and $C_6$, while the long-distance contributions
involve the operators ${\cal O}_3$,  ${\cal O}_4$, ${\cal O}_5$ and ${\cal O}_6$. Figure~\ref{fig:omom} shows the lowest-order diagrams for these
decays. New physics loops depicted in Fig.~\ref{fig:np} may also contribute here. These charmless vector
vector decays involve three amplitudes. In the transversity frame, these are the longitudinal
amplitude $A_0$ (S-wave), the transverse amplitude $A_T$ (P-wave) and the parallel amplitude $A_{||}$ 
(D-wave). For measuring \CP\ violation, they need to be known.

In chapter~\ref{chap:bsg}, we present new \babar\ measurements of the direct \CP\ asymmetry in \bsg\ 
using a semi-inclusive analysis. We extract the ratio of Wilson coefficients \imces\ from a measurement of the difference in \CP\ asymmetries between charged and neutral $B$ decays.  We also show \CP\ asymmetry measurements for \bsdg\ decays. 
In chapter~\ref{chap:bsll}, we present our branching fraction and \CP\ asymmetry measurements of \bsll\ decays using a semi-inclusive analysis. In
chapter~\ref{chap:pill}, we summarize our branching fraction upper limits on \bpll\ and \bell.  In
chapter~\ref{chap:LNV}, we summarize our results on searches for lepton number violation in exclusive
\bhll\ modes.  In chapter~\ref{chap:bomom}, we
present our results on the charmless vector vector decays \bomom\ and \bomfi\ and in chapter~\ref{chap:conclusion}
we end with concluding remarks. Note that \babar\ performs all analyses blinded meaning that
results are sealed until selection criteria and fitting procedures are finalized.
\section{Measurement of \CP\ Violation in \bsg}
\label{chap:bsg}

In the SM, the \bsg\ branching fraction is calculated at next-to-next-to-leading order (up to four loops) yielding 
${\cal B}(B \ra X_s \gamma) =(3.14 \pm 0.22) \times 10^{-4}$ for photon energies $E^*_\gamma > 1.6$~\gev\ in the
center-of-mass (CM) frame~\cite{Misiak07, Misiak07a}. For larger values of \Eg, the prediction depends the shape of the
of the \Eg\ spectrum, which is modeled in terms of a shape function~\cite{Neubert} that depends on the
Fermi motion of the $b$ quark inside the $B$ meson and thus on the $b$ quark mass. Since the shape
function is expected to be similar to that determining the lepton-energy spectrum in $B \ra  X_u \ell \nu$,
precision measurements of the \Eg\ spectrum help to determine $|V_{ub}|$ more precisely~\cite{Lange, Bauer, Gambino}. The
measurement of ${\cal B} (B\ra X_s \gamma)$ also provides constraints on the charged Higgs mass~\cite{Olive, Eigen}.

Experimentally, the challenge consists of extracting \bsg\ signal photons from those of $\pi^0$ and $\eta$ decays, 
copiously produced in $q \bar q$ continuum (with $q = u,~ d,~ s,~ c$) and
$B \bar B$ processes that increase exponentially at smaller photon energies. One strategy consists of summing $b \ra s \gamma$
exclusive final states. In a sample of $471\times 10^{-6} ~ B \bar B$ events collected with the
\babar\ detector~\cite{babar02, babar13} at the PEP-II asymmetric storage ring at the SLAC National Laboratory,
we reconstruct 38 exclusive final states containing one or three kaons with at most one $K^0_S$,
up to four pions with at most two $\pi^0$s and up to one $\eta$. We require photon energies in the CM frame of $1.6 < E^*_\gamma < 3.0$~\gev. Previously, we published total and partial branching fractions~\cite{babar12}.
Here, we focus on the measurement of direct \CP\ asymmetry, which is defined by

\begin{equation}
{\cal A}_{C\!P} (B\ra X_s \gamma) \equiv \frac{{\cal B} (\bar B \ra \bar X_s \gamma) - {\cal B} (B \ra X_s \gamma) } {{\cal B}(\bar B \ra \bar X_s \gamma) + {\cal B} (B \ra X_s \gamma)}.
\end{equation}

\begin{figure}[h]
\centering
\includegraphics[width=110mm]{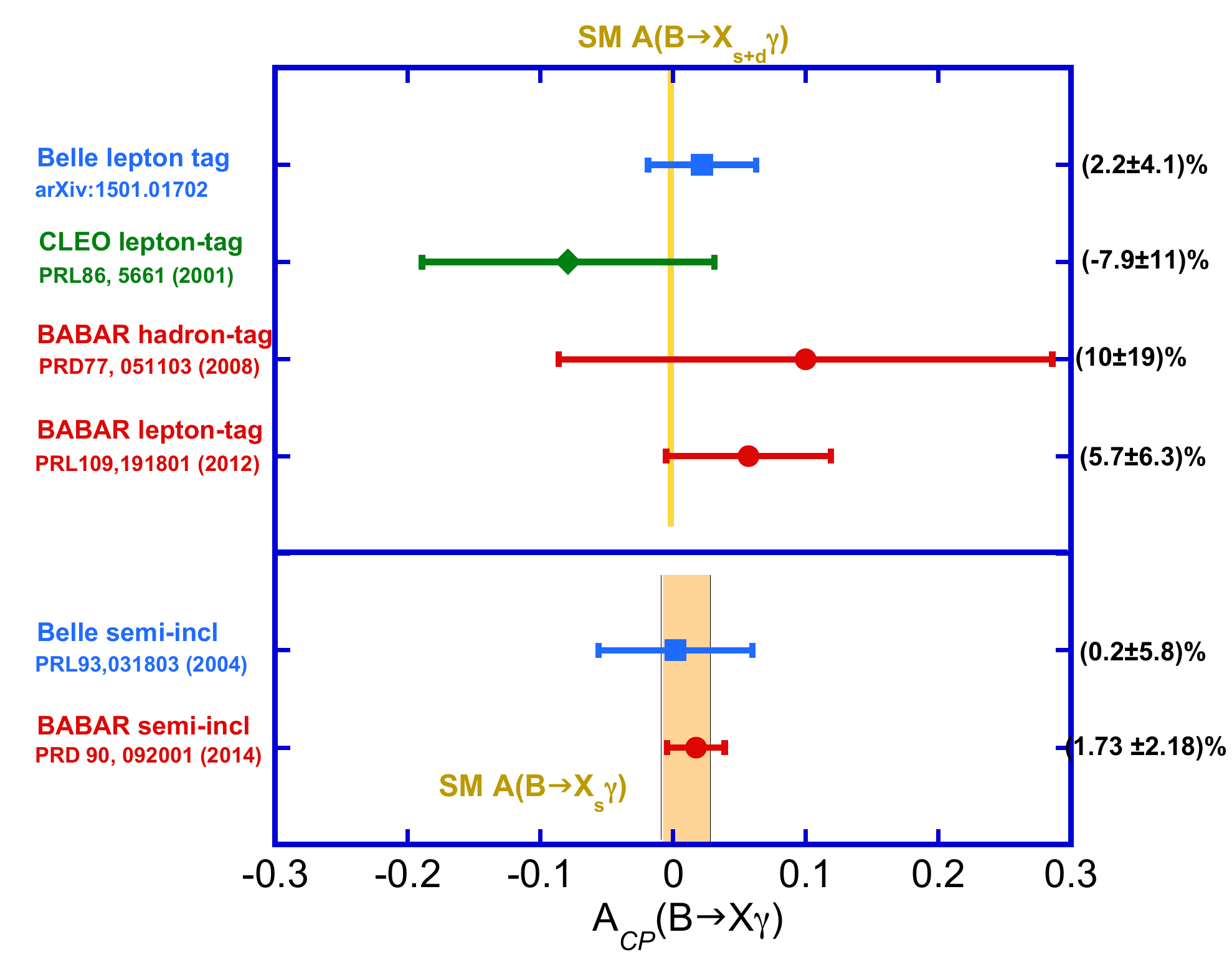}
%\vskip -0.2cm
\caption{Summary of  \acp\ measurements for \bsg\ from semi-inclusive analyses (bottom part) from \babar~\cite{babar14} and Belle~\cite{Belle04} and for $B \ra X_{s+d} \gamma$ from fully inclusive analyses (top part) from \babar~\cite{babar12a, babar12b, babar08a}, Belle~\cite{Belle14} and CLEO~\cite{cleo01}  in comparison to the SM prediction for \bsg~\cite{Benzke}  and for \bsdg~\cite{Kagan, Hurth}, respectively.}
 \label{fig:acpsg}
\end{figure}

For this analysis~\cite{babar14}, we select 16 self-tagging modes, ten $B^+ \ra h^+ \gamma$\footnote{$h^+ =
K^0_S \pi^+,~K^+\pi^0,~K^+ \pi^+ \pi^-, ~K^0_S \pi^+ \pi^0, ~K^+ \pi^0 \pi^0, ~K^0_S \pi^+ \pi^- \pi^+, 
~K^+\pi^+\pi^-\pi^0,~K^0_S \pi^+ \pi^0 \pi^0, ~K^+ \eta, ~K^+ K^+ K^-$.} and six $B^0 \ra h^0 \gamma$ final states\footnote{$h^0 = 
K^+ \pi^-, ~K^+ \pi^-\pi^0, ~K^+ \pi^+ \pi^- \pi^-,~K^+ \pi^- \pi^0 \pi^0, ~K^+\pi^-\eta, ~K^+ K^+ K^- \pi^-$.}. We maximize the signal extraction using a bagged decision tree with six input variables.
This improves the efficiency considerably with respect to the standard $\Delta E = E^*_B - E^*_{beam}$ selection where $E^*_{beam}$ and $E^*_B$
are the beam energy and $B$ meson energy in the CM frame, respectively. To remove continuum background, we train a separate bagged decision tree using event shape variables. For each $X_s$ mass bin, we optimize the sensitivity $S/\sqrt{(S + B)}$ where
$S (B)$ is the signal (background) yield using loosely identified pions and kaons. To extract
\acp, we fit the beam-energy-constrained mass $m_{ES} =\sqrt{E^{*2}_{beam}- p^{*2}_B}$ simultaneously for $\bar B$-tagged and $B$-tagged events where $p^*_B$ is the B momentum in the CM frame. After correcting
the raw \acp\ for detector bias determined from the $m_{ES}$ sideband below the signal region, we measure \acp(\bsg)
$= (1.73\pm 1.93_{stat}\pm  1.02_{sys})\%$~\cite{babar14}, which agrees well with the SM
prediction of $-0.6\% <$\acp$ < 2.8\%$ at 95\% confidence level (CL)~\cite{Benzke} and which supersedes the old
\babar\ measurement~\cite{babar08}. Though this result is the most precise single direct \acp\ measurement, the
uncertainty is sufficiently large to allow for new physics contributions in \Cseven. Figure~\ref{fig:acpsg} (bottom
part) shows our result~\cite{babar14} in comparison to the Belle measurement~\cite{Belle04}.
The \CP\ asymmetry difference between $B^+$ and $B^0$ decays, \dacpf,
 is very sensitive to new physics since it originates from the interference
between the electromagnetic and the chromomagnetic penguin diagrams in which the latter
enters through higher-order corrections. Calculations yield~\cite{Benzke}

\begin{equation}
\Delta {\cal A}_{C \! P}(B \ra X_s \gamma) \simeq 4 \pi^2 \alpha_s \frac{\bar \Lambda_{78}}{m_b} {\cal I}m \frac{C_8^{eff}}{C_7^{eff}} \simeq 0.12 \frac{\bar \Lambda_{78}}{100~\rm MeV}{\cal I}m \frac{C_8^{eff}}{C_7^{eff}},
\end{equation}
\noindent
where \lamse\ is the hadronic matrix element of the ${\cal O}_7 - {\cal O}_8$ interference, predicted to lie in the
range 17~\mev\ $<$ \lamse\ $<$ 190~\mev. In the SM, \dacp\ vanishes since \Cseven\ and \Ceight\
are real. However in new physics models, these Wilson coefficients may have imaginary parts
yielding non-vanishing \dacp~\cite{Kagan, Jung, Hayakawa}.
From a simultaneous fit to $B^+$ and $B^0$ modes, we measure \dacp$=(5.0\pm 3.9_{stat} \pm 1.5_{sys})\%$ from which we obtain the constraint -1.64 $<$ \imces\  $<$ 6.52 at 90\% CL. This is the first \dacp\ measurement and first  constraint on \imces.

\begin{figure}[h]
\centering
%\vskip -0.4cm
\includegraphics[width=75mm]{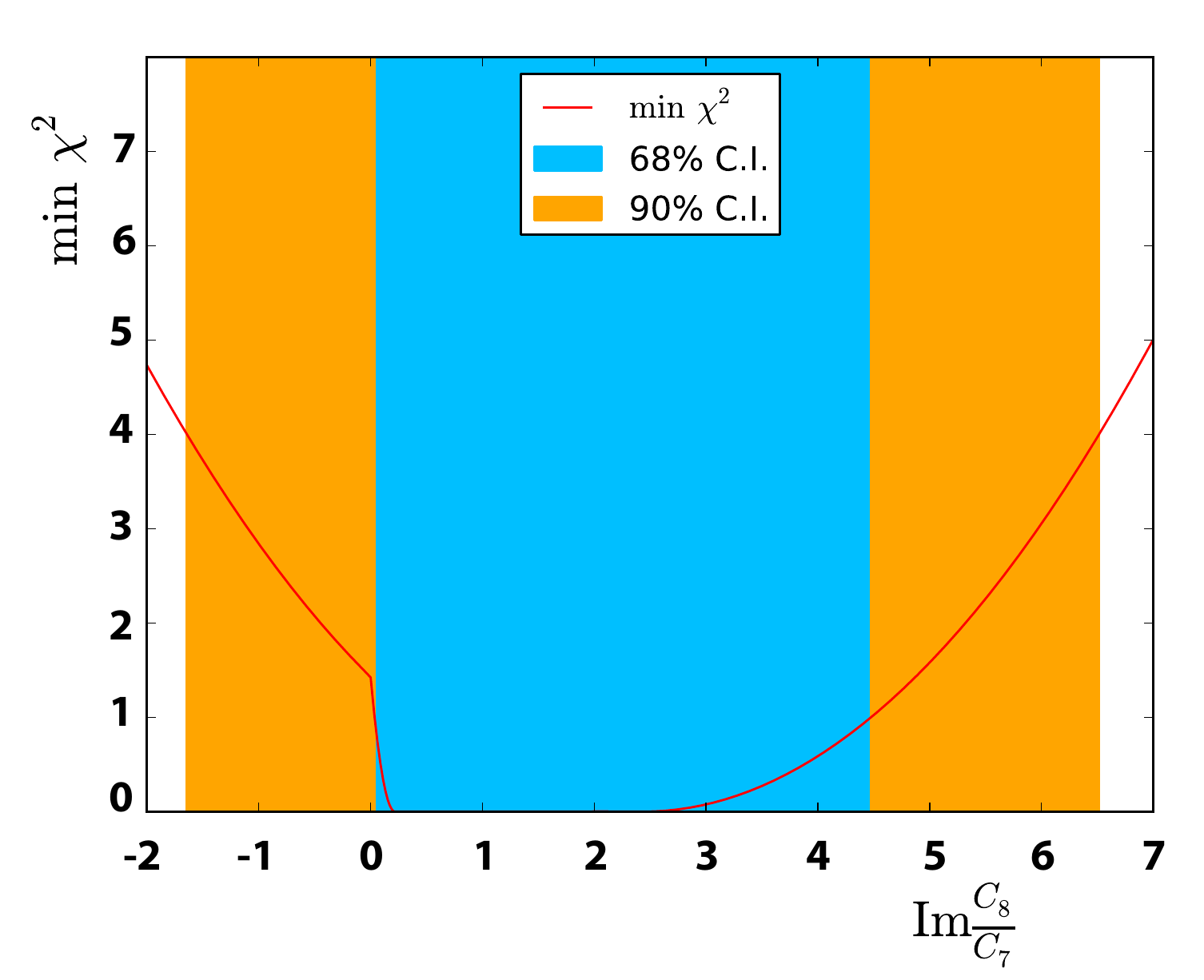}
\includegraphics[width=70mm]{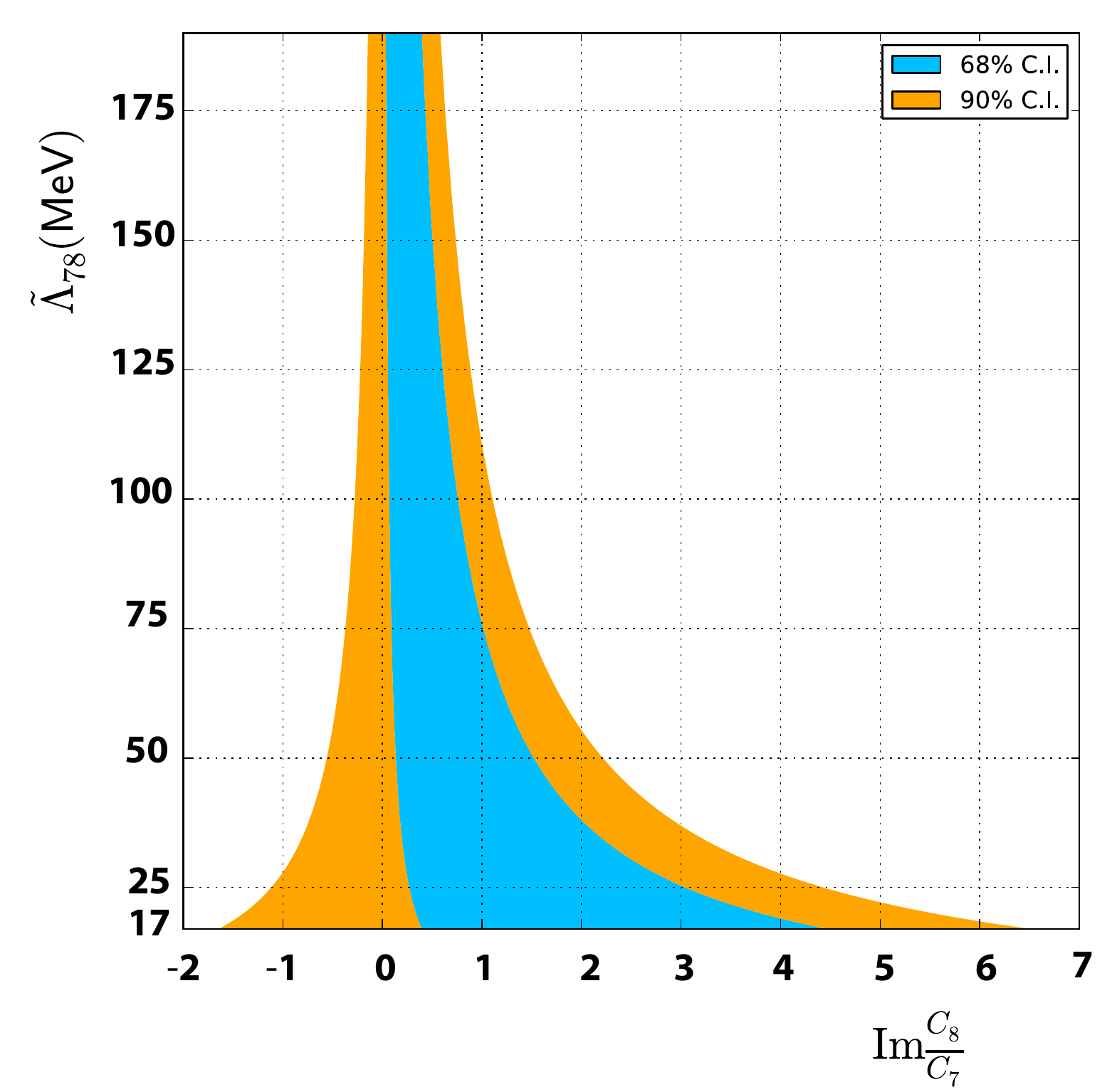}
%\vskip -0.2cm
\caption{The $\Delta \chi^2$ (left) and $\bar \Lambda_{78}$  (right) dependence on ${\cal I}m(C^{eff}_8/C^{eff}_7)$.  The blue dark-shaded (orange light-shaded) region shows the $68\%~(90\%)$ CL interval.}
 \label{fig:c78}
\end{figure}
%\vskip -0.4cm

Figure~\ref{fig:c78} (left) shows the \dchi\ of the fit as a function of \imces. The  \dchi\ dependence on \imces\ is not parabolic indicating that the likelihood has a non-Gaussian
shape. The reason is that \dchi\ is determined from all possible values of \lamse. In the region $0.2 <$\imces$ < 2.6$, a change in \imces\
can be compensated by a change in \lamse\  leaving \dchi\ unchanged. For positive values larger (smaller) than 2.6 (0.2), \dchi\ increases
slowly (rapidly) since \lamse\ remains nearly constant at the minimum value (increases rapidly).
For negative \imces\ values, \lamse\ starts to decrease again, which leads to a change in the \dchi\ shape. Figure~\ref{fig:c78} (right) shows \lamse\ as a function of \imces.

In the fully inclusive analysis, \acp\ involves contributions from \bsg\ and \bdg\ that
cannot be separated on an event-by-event basis. Therefore, we define \acp\ here as

\begin{equation}
{\cal A}_{C\! P} ( B \ra X_{s+d} \gamma) \equiv 
\frac{{\cal B}(\bar B  \ra X_{s+d} \gamma) - {\cal B} (B  \ra X_{s+d} \gamma) }
        {{\cal B}(\bar B  \ra X_{s+d} \gamma) +{ \cal B}(B  \ra X_{s+d} \gamma) }.
\end{equation}
\noindent 
We tag the flavor of the non-signal $\bar B$ flavor by the lepton charge in semileptonic decays. Using a sample $384\times 10^6~ B \bar B$
events, we measure $ {\cal A}_{C \! P} ( B \ra X_{s+d} \gamma)= 0.057 \pm 0.06_{stat} \pm  0.018_{sys}$ after correcting for charge
bias and mistagging~\cite{babar12}. Figure~\ref{fig:acpsg} (top part) shows all $ {\cal A}_{C \! P} ( B \ra X_{s+d} \gamma)$ measurements from \babar~\cite{babar12a, babar12b, babar08a}, Belle~\cite{Belle14}  and CLEO~\cite{cleo01}, which all agree well with the SM prediction~\cite{Kagan, Hurth}.

\section{Study of $ B \ra X_s \ell^+ \ell^-$ Decays}
\label{chap:bsll}

Using a semi-inclusive approach, we have updated the partial and total branching fraction measurements of \bsll\ modes ($\ell = e$ or $\mu$) with the full \babar\ data sample of $471 \times 10^6 ~ B \bar B$ events. We reconstruct 20 exclusive final states:
$K^+$, $K^0_S$, $K^{*+} (K^+ \pi^0, K^0_S \pi^+)$, $K^{*0} (K^+ \pi^-,  K^0_S \pi^0)$, $K^+ \pi^- \pi^0$, $K^0_S \pi^+ \pi^-$
with $K^0_S \ra \pi^+ \pi^-$ recoiling against $e^+ e^-$  or $\mu^+ \mu^-$~\cite{babar14a}. After accounting for $K^0_L$
modes, $K^0_S \ra \pi^0 \pi^0$ and $\pi^0$ Dalitz decays, the selected decay modes represent 70\% of the inclusive rate for hadronic masses
$m_{X_s} < 1.8$~\gevc2. Using JETSET fragmentation~\cite{JETSET} and theory predictions~\cite{Ali02, Huber, Kruger, Ali, Bobeth, Asatryan}, we
extrapolate for the missing modes and those with $m_{X_s} > 1.8$~\gevc2. We impose the requirements $m_{ES} > 5.225$~\gevc2 and $0.1~ (0.05) < \Delta E < 0.05~(0.05)$~ \gev\ for $X_s e^+e^-$ ($X_s \mu^+ \mu^-$) modes. We define six bins of the momentum-squared transferred to the dilepton system $q^2 = m^2_{\ell \ell}$ and four bins in hadronic mass $m_{X_s}$. Table~\ref{tab:bin} shows 
the defined ranges of these bins.

\vskip -0.3cm
\begin{table}[!th]
\begin{center}
\caption{Definition of the $q^2, ~ m_{\ell \ell}$ and $m_{X_s}$ bins. }
\begin{tabular}{|l|c|c||c|c|}  \hline\hline
$q^2$ bin & $q^2$  range $\rm [GeV^2/c^4]$ &  $m_{\ell \ell}$ range $\rm [GeV/c^2]$ & $m_{X_s}$ bin & $m_{X_s}$ range $ [\rm GeV/c^2] $   \B \T  \\ \hline
0 & 1.0 -- 6.0 & 1.00 -- 2.45 & &  \B \T  \\ \hline
1 & 0.1--2.0 & 0.32-- 1.41  & 1 & 0.4 -- 0.6 \B \T  \\
2 & 2.0--4.3 & 1.41--2.07   & 2 & 0.6 --1.0\B \T  \\
3 & 4.3--8.1 & 2.07 --2.6 & 3 & 1.0 --1.4\B \T  \\
4 & 10.1 --12.9 & 3.18--3.59 & 4 & 1.4 -- 1.8  \B \T  \\
5 & 14.2 -- $ (m_B - m^*_K)^2$ & 3.77 -- $ (m_B - m^*_K)$ & & \B \T  \\
 \hline\hline
\end{tabular}
\label{tab:bin}
\end{center}
\end{table}
\vskip -0.2cm

\begin{figure}[!ht]
\begin{center}
\vskip -0.3cm
\includegraphics[width=130mm]{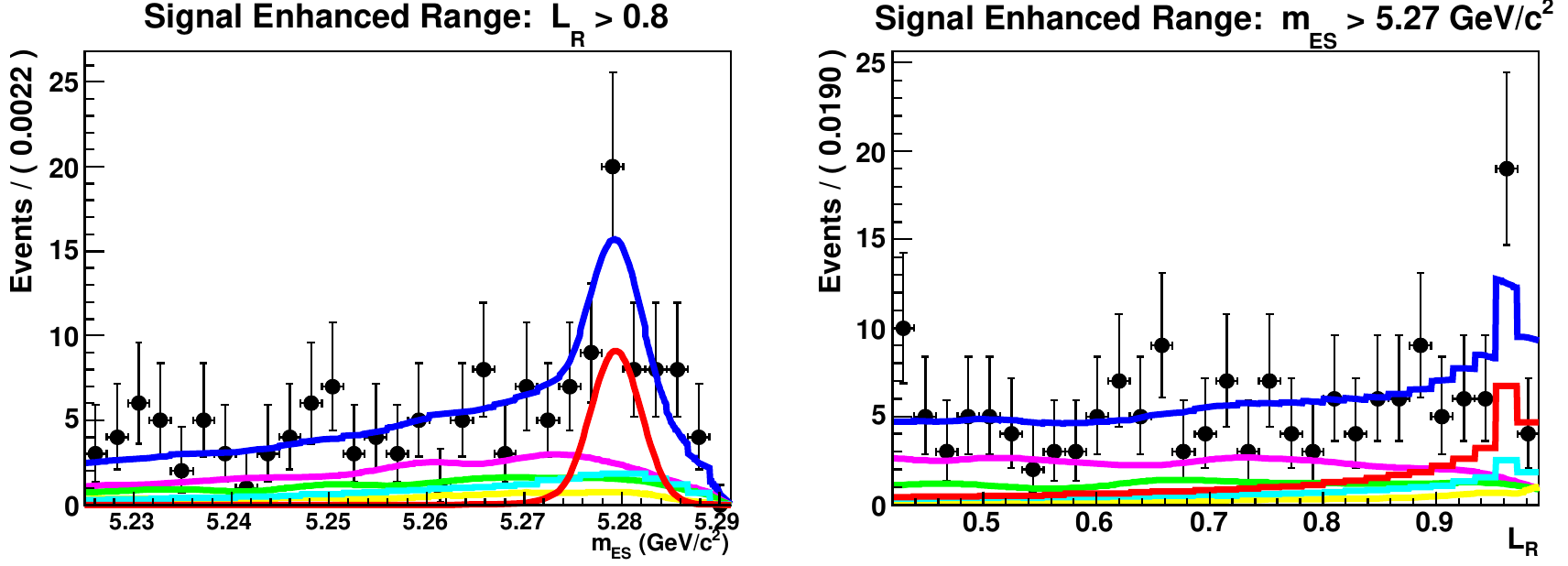}
\vskip -0.2cm
\caption{Distributions of  $m_{ES}$  (left) and likelihood ratio (right) for $B \ra X_s e^+ e^-$ in $q^2$ bin $q_5$ showing data (points with error bars), the total fit (thick solid blue curves), signal component (red peaking curves),  signal cross feed (cyan curves), $B \bar B $ background (magenta curve), $e^+ e^- \ra  q \bar q$ background (green curves) and charmonium background (yellow curves).}
\label{fig:xseebf}
\end{center}
\end{figure}

\begin{figure}[!ht]
\begin{center}
\includegraphics[width=130mm]{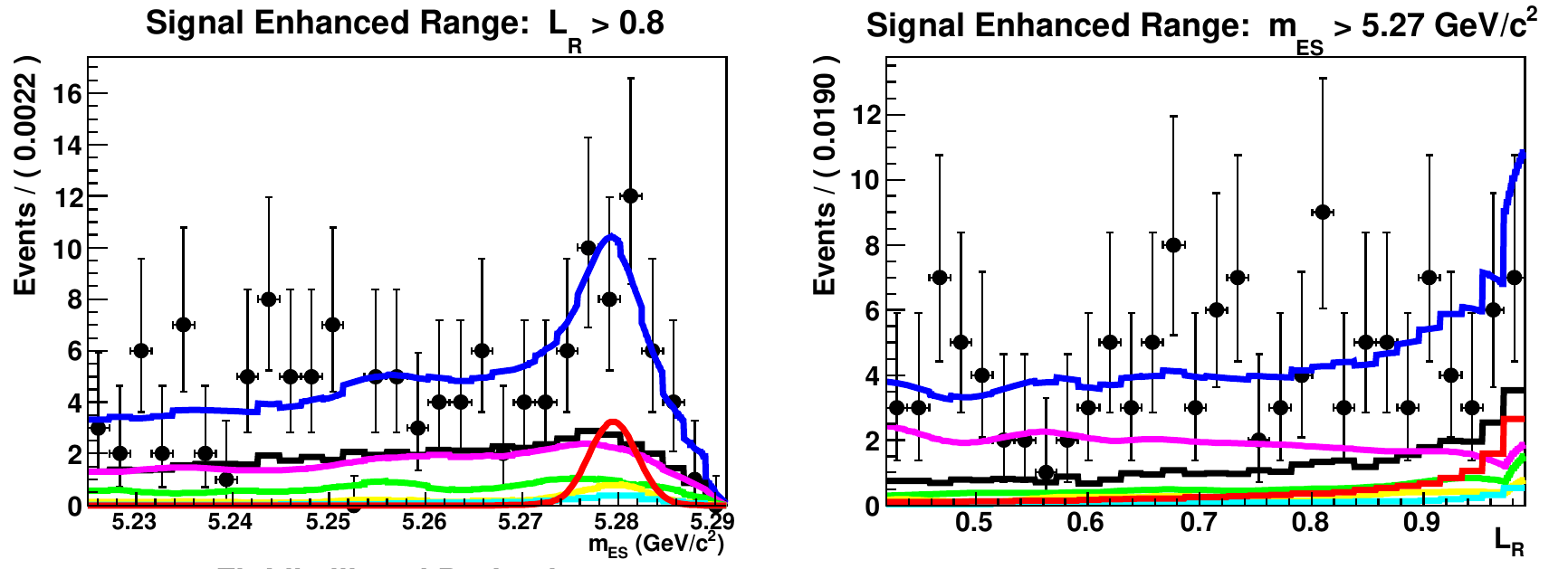}
\vskip -0.2cm
\caption{Distributions of  $m_{ES}$  (left) and likelihood ratio (right) for $B \ra X_s \mu^+ \mu^-$ in $q^2$ bin $q_1$ showing data (points with error bars), the total fit (thick solid blue curves), signal component (red peaking curves),  signal cross feed (cyan curves), $B \bar B $ background (magenta curve), $e^+ e^- \ra  q \bar q$ background (green curves) and charmonium background (yellow curves).}
\label{fig:xsmmbf}
\end{center}
\end{figure}

\begin{figure}[!ht]
\begin{center}
\includegraphics[width=75mm]{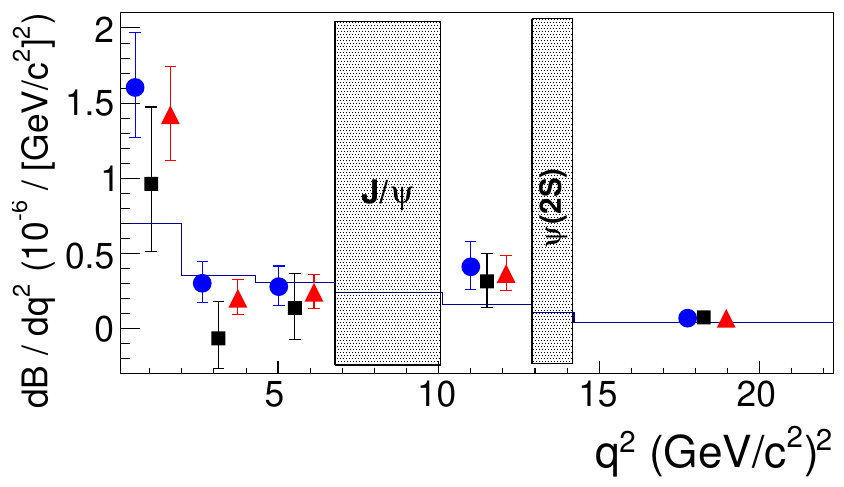}
\includegraphics[width=75mm]{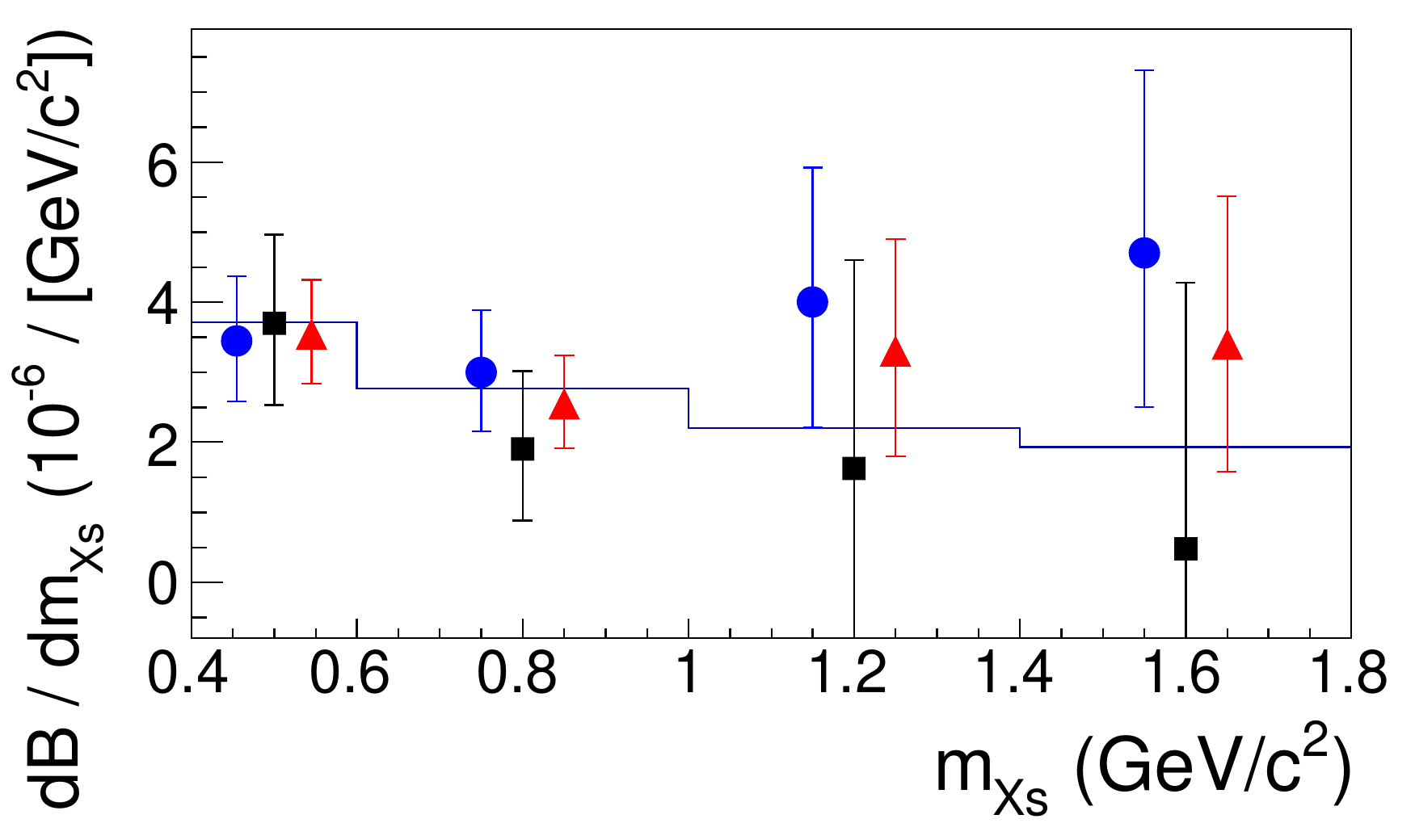}
\vskip -0.2cm
\caption{Differential branching fraction of $B \ra X_s e^+ e^-$ (blue points), $B \ra X_s \mu^+ \mu^-$ (black squares) and $B \ra X_s \ell^+ \ell^-$ (red triangles) versus $q^2$ (top) and versus $m_{X_s}$ (bottom) in comparison to the SM prediction (histogram). Grey-shaded bands show $J/\psi$ and $\psi(2S)$ vetoed regions.}
\label{fig:xsllbf}
\end{center}
\end{figure}

To suppress $e^+e^- \ra q \bar q$ and $ B \bar B$ combinatorial background, we define boosted decision trees
(BDT) for each $q^2$ bin separately for $e^+e^-$ and $\mu^+ \mu^-$ modes. From these BDTs, we determine a
likelihood ratio ($L_R$) to separate signal from $q \bar q$ and $ B \bar B$ backgrounds. We veto $J/\psi$   and  $\psi (2S)$
mass regions and use them as control samples. We measure $d{\cal B}(B \ra X_s\ell^+ \ell^- / d q^2$ in six bins of $q^2$ and four bins of $m_{X_s}$ . We extract the signal in each bin from a two-dimensional fit to $m_{ES}$ and $L_R$. As examples, Figs.~\ref{fig:xseebf} and \ref{fig:xsmmbf}
show the $m_{ES}$ and $L_R$ distributions for $e^+e^-$modes in bin $q_5$ and for $\mu^+ \mu^-$ modes in bin $q_1$,
respectively. Clear signals are visible both in the $m_{ES}$ and $L_R$ distributions.
Figure~\ref{fig:xsllbf} shows the differential branching faction as a function of $q^2$ (left) and $m_{X_s}$ (right)~\cite{babar14a}.
Table~\ref{tab:xsll3} summarizes the differential branching fractions in the low and high $q^2$ regions in
comparison to the SM predictions~\cite{Huber, Asatryan, Asatryan02, Ghinculov, Gambino03, Ghinculov03, Bobeth04, Ghinculov04, Greub, Huber08, Beneke}. In both $q^2$ regions, the differential branching fractions are in good agreement with the SM predictions. These results supersede the previous \babar\ measurements~\cite{babar04} and agree well with the measurements from Belle~\cite{Belle05}.

The direct \CP\ asymmetry is defined by

\begin{equation}
{\cal A}_{C \! P}=\frac{{\cal B}(\bar B \ra \bar X_s \ell^+ \ell^-)-{\cal B}(B \ra  X_s \ell^+ \ell^-)}{{\cal B}(\bar B \ra \bar X_s \ell^+ \ell^-)+{\cal B}(B \ra  X_s \ell^+ \ell^-)}.
\end{equation}
\vskip -0.2 cm

\begin{table}[!th]
\begin{center}
\caption{Measured $B \ra X_s \ell^+ \ell^-$ branching fractions in the low and high $q^2$ regions from \babar~\cite{babar14a} and the SM predictions. Uncertainties are statistical, systematic and from model dependence, respectively.}
\vskip 0.2 cm
\begin{tabular}{|l|c|c||c|c|}  \hline\hline
Mode & \babar\  $[10^{-6}]$ & SM $[10^{-6}]$  & \babar\  $[10^{-6}]$ & SM $[10^{-6}]$  \B \T  \\ \hline
$q^2  [\rm GeV^2/c^4 ]$ & 1  -- 6   & 1 -- 6  &  $ >14.2 $  & $ >14.2~  $  \B \T  \\ \hline
$B \ra X_s \mu^+ \mu^-$ & $0.66^{+0.82+0.30}_{-0.76-0.24}\pm 0.07 $ & $1.59\pm 0.11$ &  $0.60^{+0.31+0.05}_{-0.29-0.04}\pm 0.00 $ &  $0.25^{+0.07}_{-0.06}$  \B \T  \\
$B \ra X_s e^+ e^-$ & $1.93^{+0.47+0.21}_{-0.45-0.16}\pm 0.18 $ & $1.64\pm 0.11$  &  $0.56^{+0.19+0.03}_{-0.18-0.03}\pm 0.00$ &$0.25^{+0.07}_{-0.06}$  \B \T  \\
$B \ra X_s \ell^+ \ell^-$ & $1.60^{+0.41+0.17}_{-0.39-0.13}\pm 0.07$ & &  $0.57^{+0.16+0.03}_{-0.15-0.02}\pm 0.00$ &  $0.25^{+0.07}_{-0.06}$  \B \T  \\ \hline
 \hline\hline
\end{tabular}
\label{tab:xsll3}
\end{center}
\end{table}

\begin{figure}[!ht]
\begin{center}
\includegraphics[width=87mm] {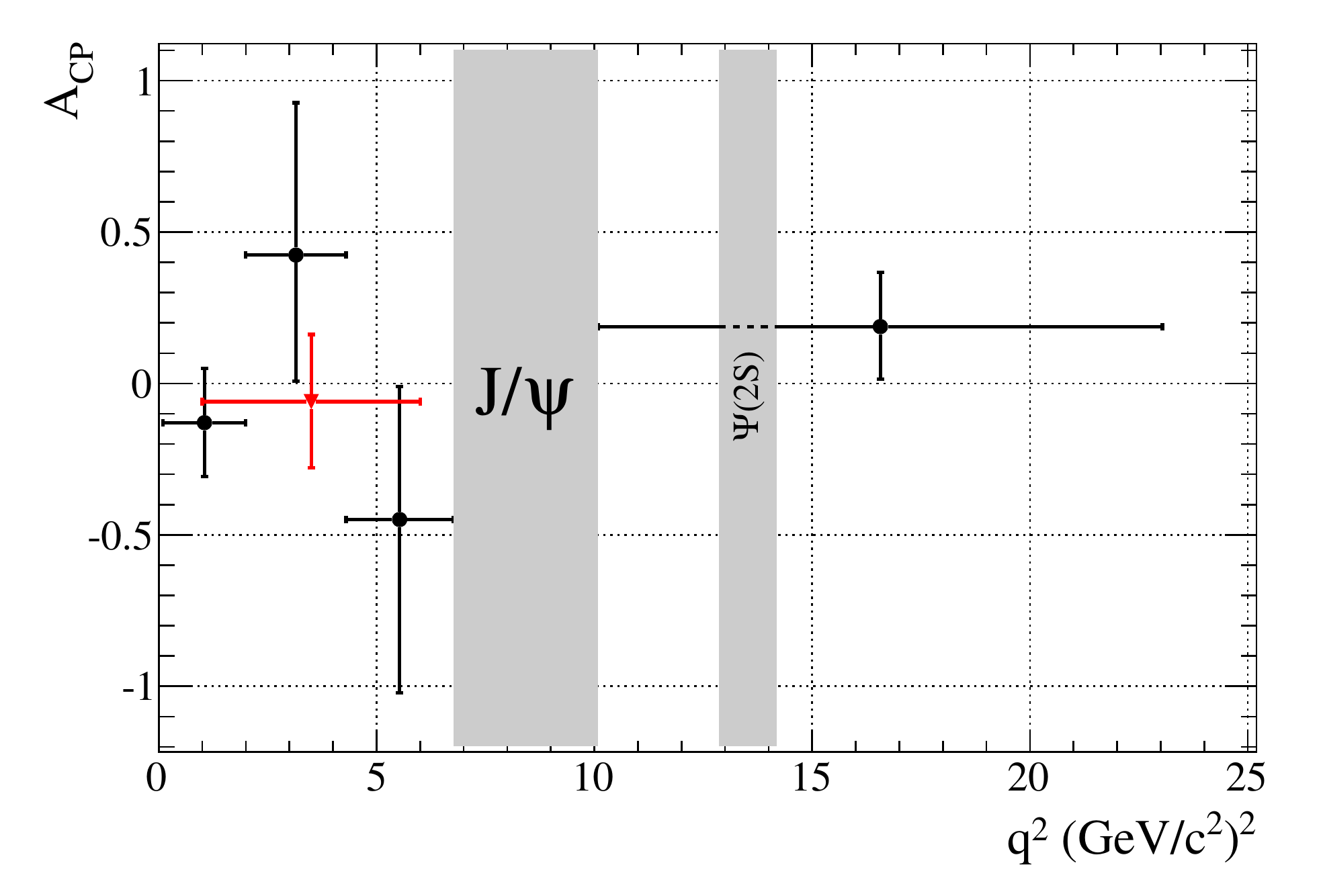}
\vskip -0.3cm
\caption{The \CP\ asymmetry versus $q^2$. Grey bands show the $J/\psi$ and $\psi(2S)$ vetoed regions.}
\label{fig:xsllcp}
\end{center}
\end{figure}
\vskip -0.3cm

We use 14 self-tagging modes consisting of all $B^+$ modes and $B^0$ modes with decays to a
$K^+$ to measure \acp$(B \ra X_s \ell^+ \ell^-)$ in five $q^2$ bins. Due to low statistics, we have combined bins $q_4$ and
$q_5$. Figure~\ref{fig:xsllcp} shows the \CP\ asymmetry as a function of $q^2$. The SM
prediction of the \CP\ asymmetry in the entire $q^2$ region is close to zero~\cite{Du, Ali99, Bobeth08, Altmannshofer}. In new physics
models, however, \acp\ may be significantly enhanced~\cite{Soni, Alok}. In the full range of $q^2$, we measure
\acp$(B \ra X_s \ell^+ \ell^-)  = 0.04 \pm 0.11_{stat}\pm 0.01_{sys}$~\cite{babar14a}, which is in good agreement with the SM prediction. The
\CP\ asymmetries in the five $q^2$ bins are also consistent with zero.

\section{Search for \bpell\ Decays}
\label{chap:pill}

In the SM in lowest order, \bdll\ modes are also mediated by the electromagnetic penguin,
$Z$ penguin and $WW$ box diagrams. However, they are suppressed by $|V_{td}/ V_{ts}|^2 \simeq 0.04$ with
respect to the corresponding \bsll\ decays. In extensions of the SM, rates may increase significantly~\cite{Aliev}.
Using $471\times 10^6~ B \bar B$ events, we recently updated the search for \bpll\ modes and performed the first search for \bell\ modes~\cite{babar13a}. The SM predictions lie in the range 
${\cal B}(B \ra \pi \ell^+ \ell^-) = (1.96$--$3.30) \times 10^{-8}$ and ${\cal B}(B \ra \eta \ell^+ \ell^-) = (2.5$--$3.7) \times 10^{-8}$ where the
large ranges result from uncertainties in the $B \ra \pi$ form factor calculations~\cite{Aliev, Wang, Song}
and from a lack of knowledge of $B \ra \eta$ form factors~\cite{Erkol}, respectively.

We fully reconstruct four \bpll\ and four \bell\ final states by selecting $\pi^\pm,~ \pi^0 \ra \gamma \gamma$ and
$\eta \ra \gamma \gamma, ~\pi^+ \pi^- \pi^0$  recoiling against $e^+e^-$ or $\mu^+ \mu^-$. We select leptons with $p_\ell  > 0.3~ \rm GeV/c$, recover  $e^\pm$ bremsstrahlung losses, remove $\gamma$ conversions and require good particle identification for $e^\pm, ~\mu^\pm$ and $\pi^\pm$. We select photons with $E_\gamma > 50$~\mev\ and impose $\pi^0$ and $\eta$ mass  constraints of $115< m_{\gamma \gamma} < 150$~\mevc2\
and $ 500~ (535) < m_{\gamma \gamma} ~(m_{3\pi}) < 575~ (565)$~\mevc2, respectively. For the $\eta \ra \gamma \gamma$ final state, we require $(E_{1,\gamma} - E_{2,\gamma} )/(E_{1,\gamma} + E_{2,\gamma} ) < 0.8$ to remove asymmetric \qq\ background that peaks near one. We
veto \jpsi\ and \psip\ mass regions and use four neural networks (NN) to suppress combinatorial \BB\ and \qq\
continuum backgrounds, separately for \ee\ and for \mm\ modes. The NNs for suppressing \BB\
background use 15 (14) input distributions for \ee\ (\mm) modes, while those for suppressing \qq\ continuum use 16 input distributions for both modes. For validations of the fitting procedure and peaking backgrounds, we use pseudo-experiments and the vetoed \jpsi\  and \psip\ samples.

For $B^+ \ra \pi^+ \ell^+ \ell^-$  and $B^0 \ra \pi^0 \ell^+ \ell^-$, we perform simultaneous unbinned maximum likelihood (ML) fits to \mes\ and \DE\ 
distributions for \ee\ and \mm\ modes separately. We include the $B^+ \ra K^+  \ell^+ \ell^-$ mode in the fit to extract the peaking background contribution in the $B^+ \ra \pi^+ \ell^+ \ell^-$ modes by reconstructing the $K^+$  as a $\pi^+$. For $B^0 \ra \eta \ell^+ \ell^-$, we perform
simultaneous unbinned ML fits to \mes\ and \DE\  distributions, again for \ee\ and \mm\ modes separately. In addition, we perform fits for the isospin-averaged modes $B \ra \pi e^+ e^-$ and $B \ra \pi \mu^+ \mu^-$, lepton-flavor-averaged modes $B^+ \ra \pi^+ \ell^+ \ell^-$, 
$B^0 \ra \pi^0 \ell^+ \ell^-$ and $B^0 \ra \eta \ell^+ \ell^-$ and both isospin- and lepton-flavor-averaged modes $B \ra \pi \ell^+ \ell^-$.

\begin{figure}[h]
\centering
\includegraphics[width=100mm]{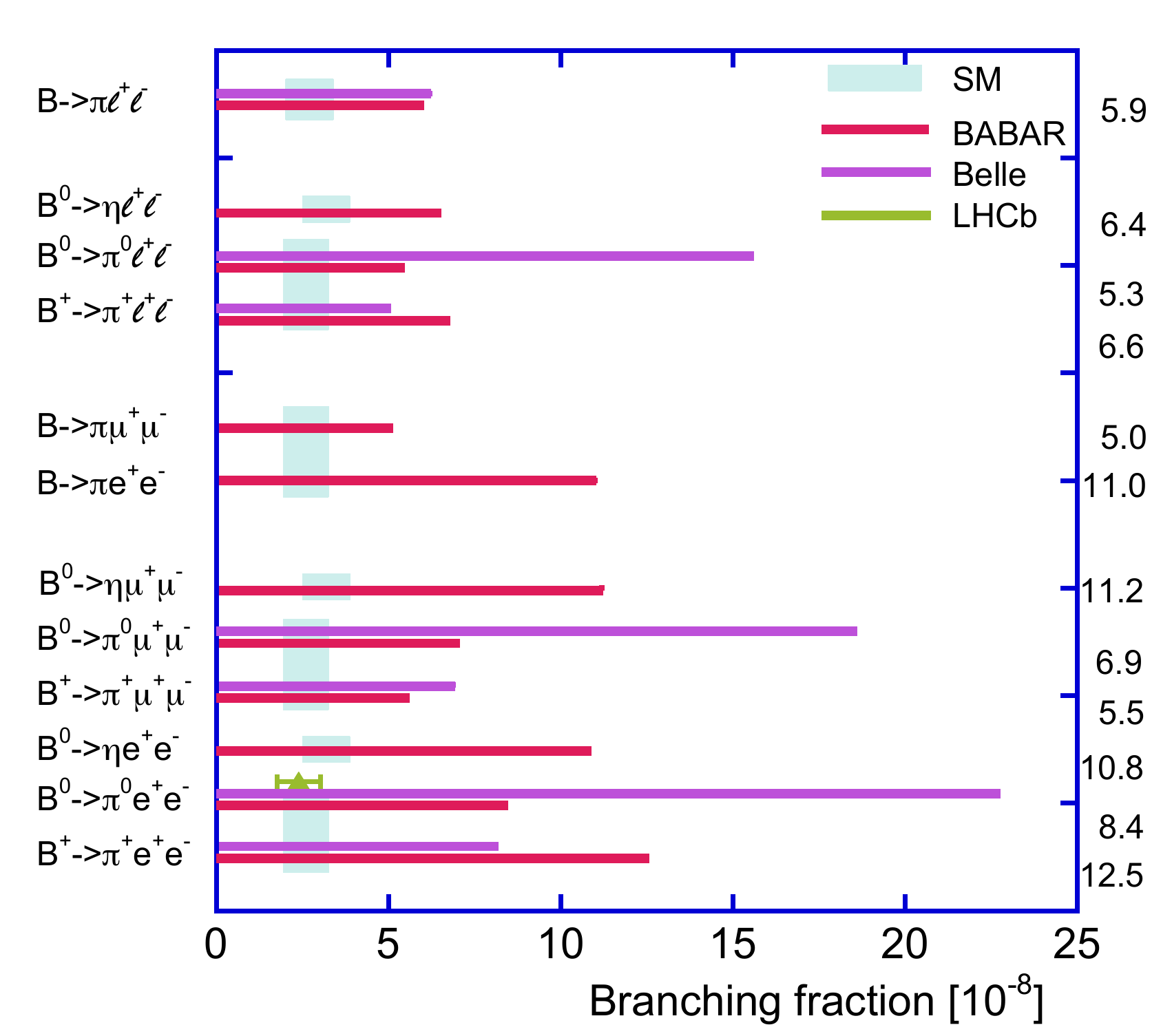}
\caption{Branching fraction upper limits at 90\% CL for \bpll\ and \bell\ modes from \babar~\cite{babar13a} 
and Belle~\cite{Belle08a} and the measurement of  $B^+ \ra \pi^+ \ell^+ \ell^-$ from LHCb~\cite{lhcb12c}.}
\label{fig:dll}
\end{figure}

We see no signals in any of these modes and set branching fraction upper limits at 90\%
CL. Figure~\ref{fig:dll} shows them in comparison to results from Belle~\cite{Belle08a}  and a measurement of
${\cal B} (B^+ \ra \pi^+ \mu^+ \mu^-)$  from LHCb~\cite{lhcb12c}. For $B^0 \ra \pi^0 \ell^+ \ell^-$, our branching fraction upper limit is the
lowest and so far only \babar\ has  searched for $B^0 \ra \eta \ell^+ \ell^-$ modes. The present branching fraction upper
limits lie within a factor of two to three of the SM predictions.

\section{Search for Lepton Number Violation in \bhll\ Decays }
\label{chap:LNV}

In the SM, lepton number is conserved in low-energy collisions. However, in high-energy and
high-density interactions, lepton number may be violated~\cite{Klinkhammer}. Many models beyond the SM
predict lepton number violation (LNV) with rates~\cite{Atre} that may be accessible already in present
data samples. These models also predict Majorana-type neutrinos~\cite{Majorana} for which particles and
antiparticles are identical. Via oscillation of a neutrino into an antineutrino, lepton-number
violating decays become possible such as $B^+ \ra K^- \ell^+ \ell^{\prime -}$ depicted in Fig.~\ref{fig:np} (right). The
observation of atmospheric neutrino oscillations confirms hat neutrinos carry mass~\cite{Fukuda} but we
do not know if any Majorana-type neutrinos exist. However, lepton number violation is also a
necessary condition to explain the observed baryon asymmetry in the universe~\cite{Davidson}.

Using the full \babar\ data set of $471\times 10^6~ B \bar B$ events collected at the $\Upsilon(4S)$ peak, we
have searched for lepton number violation in 11 $B^+$ decays\footnote{$B^+ \ra \rho^- (\pi^- \pi^0) \ell^+ \ell^{\prime +},~B^+ \ra K^{*-} (K^0_S \pi^-$ and $K^- \pi^0),  ) \ell^+ \ell^{\prime +},~B^+ \ra D^- (K^- \pi^+ \pi^- ) \ell^+ \ell^{\prime +},~B^+ \ra \pi^- e^+ \mu^+$ and $B^+ \ra \pi^- e^+ \mu^+$ where $\ell^+, \ell^{\prime +} = e^+$ or $\mu^+$.}~\cite{babar14b}. We select events with more
than three charged tracks of which two are identified as like-sign leptons having a combined
momentum less than $2.5~\rm GeV/c$ in the laboratory frame. We remove $e^+$ and $e^-$ from photon
conversions. We define sufficiently wide mass regions around the $\rho^- ~( 0.47 < m_{\pi \pi^0} < 1.07$~\gevc2),
$K^{*-}~(0.770 < m_{K \pi} < 1.01$~\gevc2) and $D^- ~(1.835 < m_{K \pi \pi} < 1.895$~\gevc2) mesons to allow reasonable
modeling of backgrounds. We combine the $\pi^-, ~ K^-, ~\pi^- \pi^0,~ K^- \pi^0,~K^0_S \pi^-$ and $ K^- \pi^+ \pi^-$ candidates with the two
leptons to form a $ B$ candidate. We remove $\ell^\pm h^\mp$ combinations with an invariant mass close to
that of the \jpsi\ as the $h^\mp$ may be a misidentified $\mu^\mp$ from $J/\psi \ra \mu^+ \mu^-$. The misidentification 
rate is about 2\%.

\begin{table}[!th]
\begin{center}
\caption{Summary of signal yield after fit bias correction with statistical uncertainty, reconstruction efficiency $\epsilon$, significance S including systematic error, measured branching fraction ${\cal B} \times 10^7$ and the 90\% CL upper limit 
${\cal B}_{UL} \times 10^7$. Yields and efficiencies for the \bkstll\ modes are for $K^- \pi^0$ and $K^0_S \pi^-$ final states, respectively.}
\vskip 0.2 cm
\begin{tabular}{|l|c|c|c|c|c|}  \hline\hline
Mode & Yield &  $\epsilon$ & $S$ & ${\cal B}  $&  ${\cal B}_{UL}$  \B \T  \\ \hline
 & [events] &  $ [\%]$ & $[\sigma]$ & $[10^{-7}] $&  $[ 10^{-7}] $  \B \T  \\ \hline
$K^{*-} e^+ e^+$   & $3.8\pm 3.3,~ 0.8\pm 3.9$  & $11.5\pm 0.1,~ 12.1\pm 0.1$ & $1.2 $&   $1.7\pm 1.4\pm 0.1 $&  $4.0 $ \B \T  \\ \hline
$K^{*-} e^+ \mu^+$ &  $-1.9\pm 4.7,~ -5.1\pm 2.6$  & $7.9\pm 0.1,~ 8.5\pm 0.1$ &  $ 0.0$&  $ -4.5\pm2.6\pm0.4$&  $3.0 $ \B \T  \\
$K^{*-} \mu^+ \mu^+$ &    $2.3\pm 1.8,~ 2.0\pm 1.8$  & $6.1\pm 0.1, ~5.8\pm 0.1$ & $ 1.3$&  $ 2.4\pm1.8\pm0.4$&  $5.9 $ \B \T  \\ \hline
$\rho^- e^+ e^+$  & $-2.1\pm 5.7$  & $12.1\pm 0.1$ &  $0.0 $&  $-0.4\pm1.0\pm0.1 $&  $1.7 $ \B \T  \\
$\rho^- e^+ \mu^+$  &   $4.6\pm 11.4$  & $10.3\pm 0.1$ &  $0.4 $&  $1.0\pm2.4\pm0.2 $&   $ 4.7$ \B \T  \\
$\rho^- \mu^+ \mu^+$  &   $2.9\pm 6.8$   & $7.3\pm 0.1$ &  $ 0.5$&  $ 0.9\pm2.0\pm0.3$&  $4.2 $ \B \T  \\ \hline
$D^- e^+ e^+$  &    $3.9\pm 4.8$   & $10.2\pm 0.1$ &  $1.0 $&  $8.8\pm8.6\pm1.5 $&   $26 $ \B \T  \\
$D^- e^+ \mu^+$  &  $1.1\pm 3.2$   &  $7.7 \pm 0.1$ &  $ 0.5$&  $3.4\pm9.4\pm1.1 $&  $21 $  \B \T  \\
$D^- \mu^+ \mu^+$  & $-1.7\pm 2.5$& $5.7\pm 0.1$ &     $ 0.0$&  $-6.5\pm9.9\pm0.9 $&   $17 $ \B \T  \\ \hline
$K^- e^+ \mu^+$ &   $5.5\pm 3.5$  &  $15.2\pm 0.1$ &  $1.8 $&  $0.6\pm0.5\pm0.1 $&   $1.6 $ \B \T  \\ \hline
$\pi^- e^+ \mu^+$ & $3.8\pm 3.5$  &  $16.4\pm 0.2$ &  $1.2 $&  $0.5\pm 0.5\pm 0.1 $&   $ 1.5$ \B \T  \\
 \hline\hline
\end{tabular}
\label{tab:xllp}
\end{center}
\end{table}

For each signal mode, we construct BDTs to discriminate signal from \BB\ and \qq\ backgrounds
using nine inputs consisting of event shape variables, kinematic observables,  flavor tagging and
the proper decay time. If more than one candidate is found, we choose the one with the
smallest $\chi^2$ in the fit to the B decay vertex. We perform a simultaneous unbinned ML
fit to \mes, \DE\ and the BDT output distributions. For the \brholl, \bkstll\ and \bdmll\ final states, we include the 
$m_{\pi \pi}, ~ m_{K \pi}$ and $m_{K \pi \pi}$ mass distributions, respectively. The background PDFs consist of an Argus function~\cite{Argus} 
for \mes, first- or second-order polynomials (for $K^-,~ \pi^-,~\rho^-,~K^{*-}$ modes) or a Cruijff\footnote{The Cruijff function is a centered Gaussian with different left-right resolutions and non-Gaussian tails: $f(x) =\exp{((x-m)^2/(2\sigma^2_{L,R} + \alpha_{L,R} (x-m)^2)}$. } function 
(for $D^-$ modes) for  \DE, a non-parametric kernel estimation KEYS algorithm~\cite{KEYS} for the BDT output and a first-order 
polynomial plus a Gaussian function for the resonance masses. The corresponding signal PDFs consist of a Crystal
Ball function~\cite{Xtal} for \mes, a Crystal Ball function plus a  first-order polynomial (for modes with $\pi^0$s) for \DE, 
the simulated distribution in form of a histogram for the BDT output and for the
$D, ~K^*,~ \rho$ masses two Gaussians, a relativistic Breit-Wigner function and a Gounaris-Sakurai
function~\cite{GS}, respectively. We checked the  fit procedure with a simulated background sample having the same size as the on-resonance data sample. We further performed a blinded fit to the on-resonance data sample confirming that the background distributions agreed with the
background PDFs. Selection efficiencies vary between 6\% and 16\% depending on the final state.  

\begin{figure}[h]
\centering
\includegraphics[width=50mm]{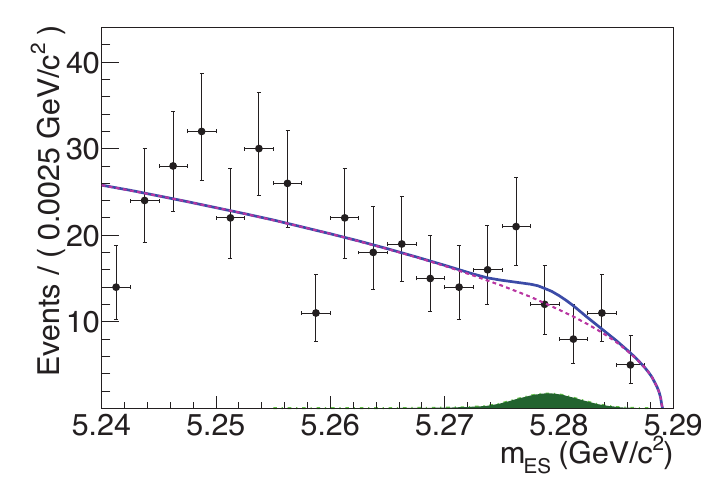}
\includegraphics[width=50mm]{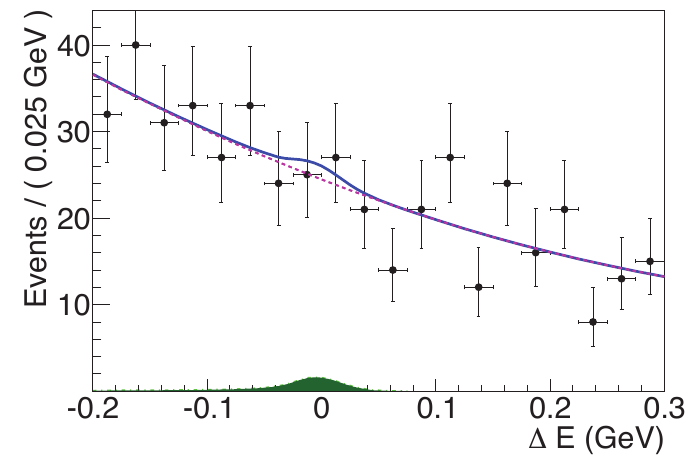}
\includegraphics[width=50mm]{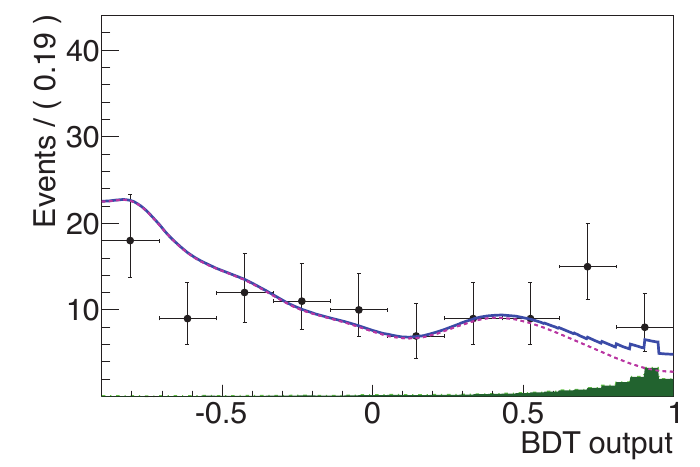}
\caption{Projections of the multidimensional fit onto \mes\ (left), \DE\ (middle) and the BDT output (right) for $B^+ \ra \pi^- e^+ \mu^+$ showing data (points with error bars), total fit (solid blue line), signal PDF (green histogram) and background (dashed magenta line).}
\label{fig:pllp}
\end{figure}

\begin{figure}[h]
\centering
\includegraphics[width=55mm]{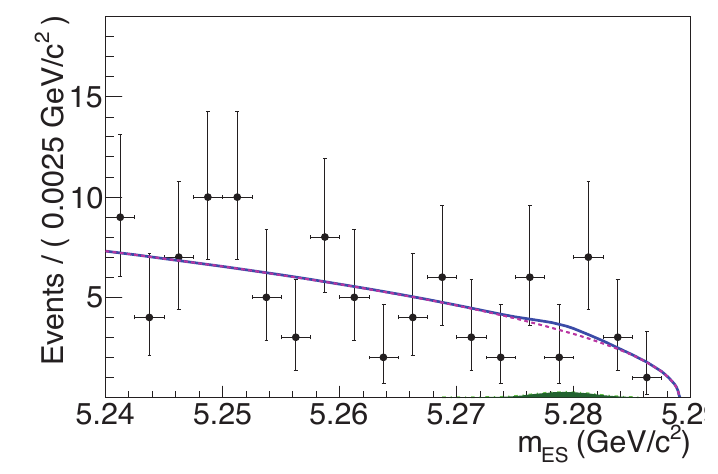}
\includegraphics[width=55mm]{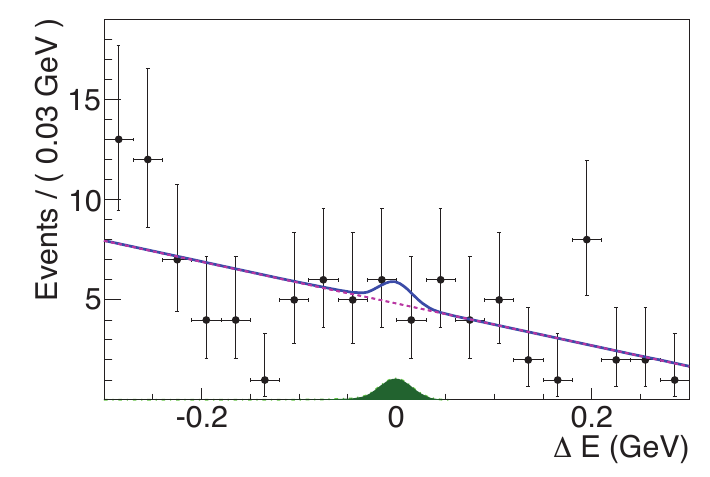}
\includegraphics[width=55mm]{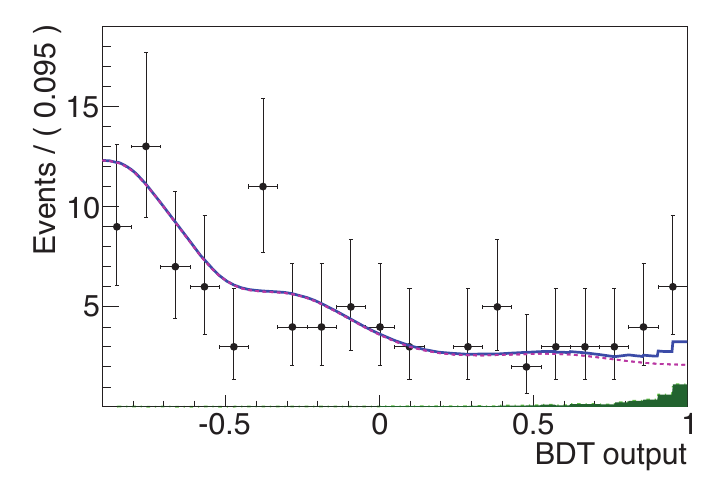}
\includegraphics[width=55mm]{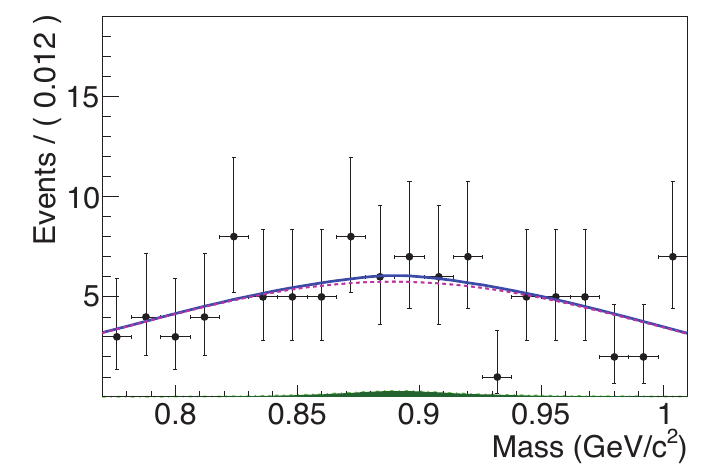}
\caption{Projections of the multidimensional fit onto \mes\ (top left), \DE\ (top right), the BDT output (bottom left) and $K^{*-}$ mass  (bottom right) for $B^+ \ra K^{*-} ~(K^0_S \pi^-) e^+ \mu^+$ showing data (points with error bars), total fit (solid blue line), signal PDF (green histogram) and background (dashed magenta line).}
\label{fig:kstllp}
\end{figure}

\begin{figure}[h]
\centering
\includegraphics[width=90mm]{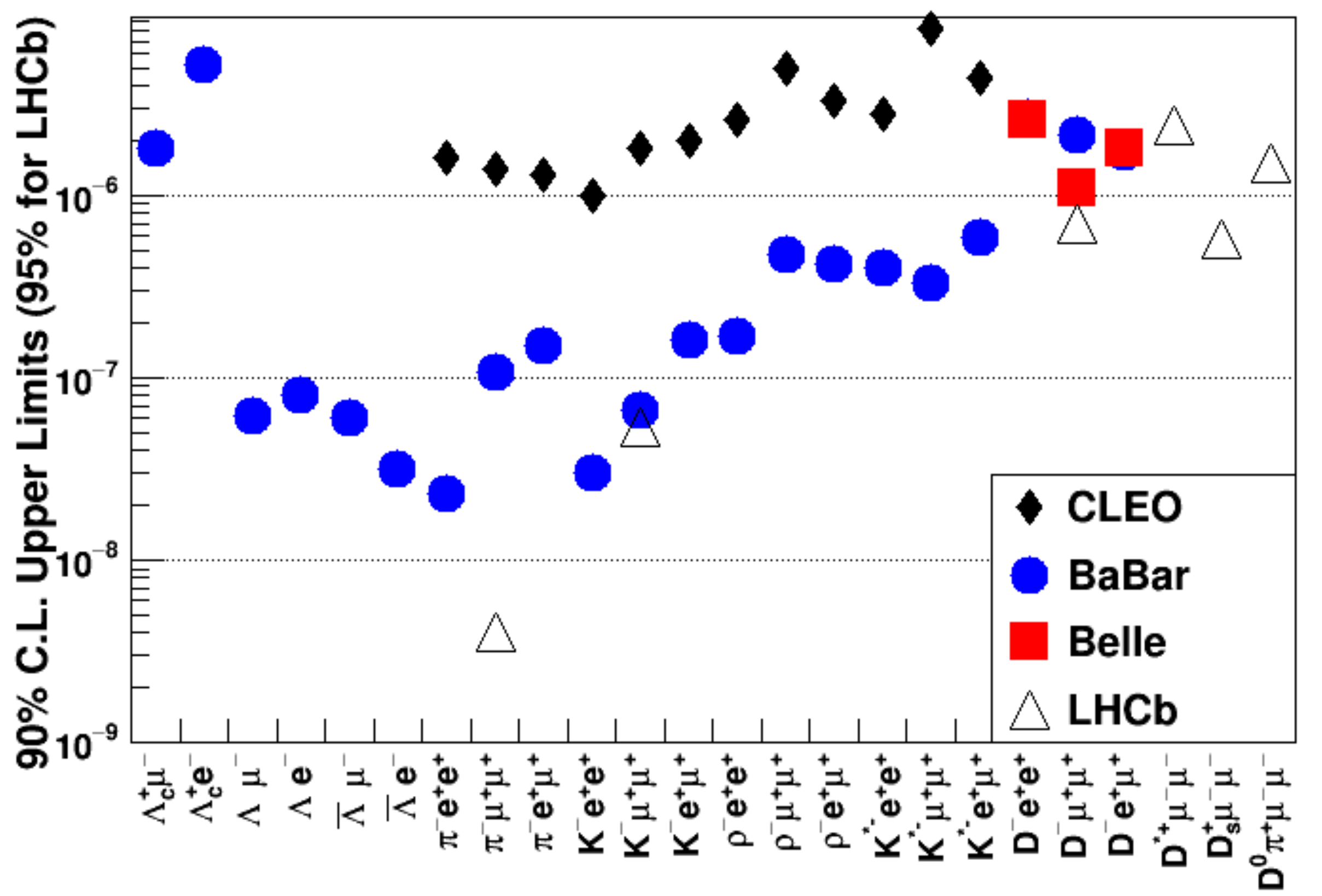}
\caption{Branching fraction upper limits at 90\% CL for LNV decays from \babar~\cite{babar14b} (solid blue points), Belle~\cite{Belle11} (solid red squares) and CLEO~\cite{cleo02} (black diamonds). In addition, LHCb upper limits at 95\% CL~\cite{lhcb14} (open triangles) are shown.}
\label{fig:LNV}
\end{figure}

Figures~\ref{fig:pllp} and \ref{fig:kstllp} show projections of the fit on the discriminating variables for $B^+ \ra \pi^- e^+ \mu^+$ and 
$B^+ \ra K^{*-} (K^0_S \pi^-) \mu^+ \mu^+$, respectively. Table~\ref{tab:xllp} summarizes our results. In all 11 modes, the data are
consistent with combinatorial background. We see the highest significance of $1.8\sigma$ in $B^+ \ra K^- e^+ \mu^+$. We set Bayesian upper limits on the branching fraction at 90\% CL using a flat prior (see Tab.~\ref{tab:xllp}). The additive systematic uncertainty that includes contributions from the PDF parameterization, fit biases, background yields and efficiencies is mode dependent
between 0.2 and 0.7 events. The total multiplicative uncertainty on the branching fraction is
5\% or less. The branching fraction upper limits at 90\% CL lie in the range 1.5 --  26 $\times 10^{-7}$
where the lowest limit is set in the $B^+\ra  \pi^-  e^+ \mu^+$ mode. Figure~\ref{fig:LNV} summarizes all results
of lepton-number-violating B decays from \babar~\cite{babar14b}, Belle~\cite{Belle11}, LHCb~\cite{lhcb14} and CLEO~\cite{cleo02}
including results for $B^0\ra \Lambda^+_c \ell^-, ~ B^- \ra \Lambda \ell^-, ~ B^- \ra \bar \Lambda \ell^-$~\cite{babar11} and
 $B^- \ra \pi^-/K^-  \ell^+ \ell^+$~\cite{babar12c}. All limits are set at 90\% CL except for LHCb whose limits are set at 95\% CL.

\section{Study of \bomom\ and \bomfi\ Decays}
\label{chap:bomom}

\begin{figure}[h]
\centering
\includegraphics[width=70mm]{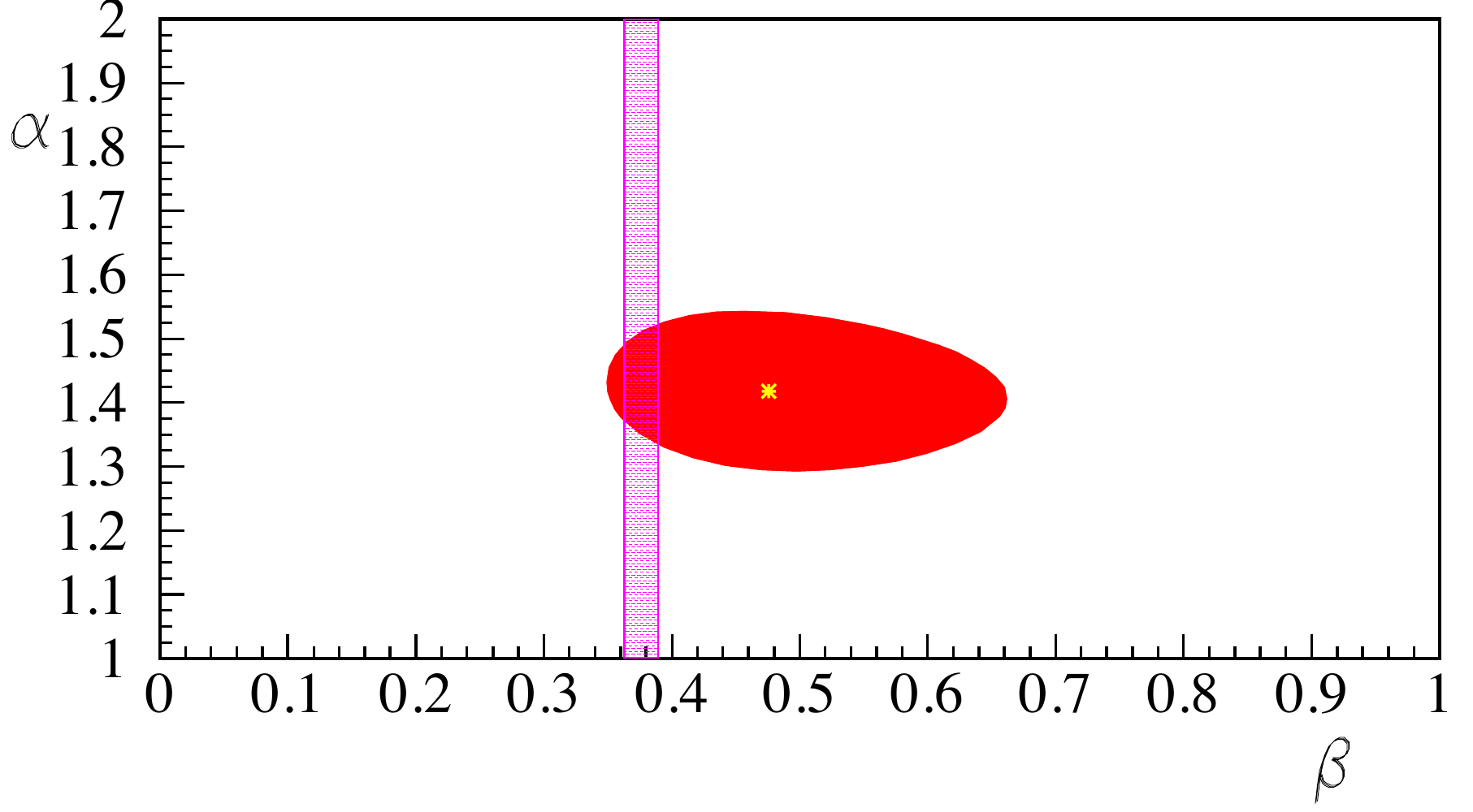}
\vskip -0.3cm
\caption{The $\alpha$--$\beta$ contour obtained from a ML fit to branching fractions, longitudinal polarizations and \CP\ asymmetries measured in all observed $B \ra V V$ modes. The magenta band shows the result of the $\sin 2 \beta$  measurements in $B \ra c \bar c K^0_{ S,L}$ decays~\cite{HFAG}.}
\label{fig:alpha}
\end{figure}

The longitudinal polarization fraction \FL\ in charmless vector vector decays poses a puzzle.
In tree-dominated decays like $B^0 \ra \rho^+ \rho^-$ and $B^+ \ra \rho^+ \rho^0$, \FL\ is nearly 100\% while in
decays with dominant penguin contributions like $B\ra K^* \rho$, \FL\ is around 50\%~\cite{HFAG}. Are
there large transverse SM contributions that reduce \FL\ \cite{Bauer04, Colangelo, Kagan04, Ladisa, Cheng, Li} or is this caused by new
physics~\cite{Cheng, Bensalem, Giri, Alvarez, Chen, Yang, Das, Giri05, Baek, Bou, Li06, Cheng09}?
Thus, it is interesting to investigate other charmless vector vector decays such as the so far not-observed modes \bomom\ and \bomfi~\cite{Oh}. In the SM, the branching fractions are expected to be of the order of 
${\cal O} (10^{-6})$ for \bomom\ and ${\cal O} (10^{-7})$ for \bomfi. The SM predicts longitudinal polarization fractions of \FL$ >0.8$ for both
modes~\cite{Li06, Cheng09}. Charmless vector vector modes are also well suited to measure the Unitarity
Triangle angle $\alpha$~\cite{Atwood, Huang}. The Scan Method group has determined  $\alpha$ -- $\beta$ contours from  a $\chi^2$ fit to
measured branching fractions, longitudinal polarizations and \CP\ asymmetries using all observed
charmless vector vector decays~\cite{eigen14}. The decay amplitudes of each mode are expressed in terms
of tree, color-suppressed tree, gluonic penguin, singlet penguin, electroweak penguin and $W$-annihilation/$W$-
exchange amplitudes. For decays involving $K^*$s, SU(3) breaking  is taken into account. All contributions
up to order ${\cal O}(\lambda^5)$ are considered where $\lambda = \sin \theta_c$ (Cabibbo angle), since the leading amplitude is already at order ${\cal O}(\lambda^3)$. Figure~\ref{fig:alpha} shows the 90\% CL $\alpha$--$\beta$ contour determined from the fit.

Using the full \babar\ data sample of $471\times 10^6~ B \bar B$ events, we reconstruct the $B$-daughter
candidates via their decays $\omega \ra \pi^+ \pi^- \pi^0$ with $\pi^0 \ra \gamma \gamma$ and $\phi \ra K^+ K^-$~\cite{babar14c}. If multiple
candidates exist, we select the one for which the $\chi^2$ probability of a fit to the two vector meson
masses is smallest. The combinatorial background from $e^+ e^- \ra q \bar q$ collisions dominates. We use
a tight selection on the angle $\theta_T$ between the thrust of the $B$ signal candidate in the $\Upsilon(4S)$ rest frame
and that of the rest of the event, requiring $|\cos\theta_T| < 0.8~ (0.9$) for \bomom\ and \bomfi\ decay modes. Furthermore, we define 
a Fisher discriminant ${\cal F}$ based on four shape and kinetic variables as inputs. We perform an extended unbinned ML fit to extract the signal and background yields from the data. We define the PDF as a product of six individual PDFs including \mes, \DE, ${\cal F}$,
masses and helicity angles of the two vector mesons and the decay angle $\psi$ between the $\pi^0$ in the dipion
rest frame and the  $\omega$ flight direction:
\begin{equation}
{\cal P}^i_j = {\cal P}_j (m^i_{ES}) \times  {\cal P}_j (\Delta E^i) \times  {\cal P}_j ({\cal F}^i) \times  {\cal P}_j (m_{V1}^i, m_{V2}^i, \theta_1^i, \theta_2^i) \times  {\cal P}_j (\psi_{\omega_1}^i) \times  {\cal P}_j (\psi_{\omega_2}^i) 
\end{equation}
\noindent
where the last term is not present in \bomfi. For signal, we use a sum of two Gaussians for \mes\ and \DE, a two-piece normal distribution for ${\cal F}$, relativistic Breit-Wigner functions for the $m_V$ distributions, each convolved with two Gaussians to account for detector resolution. We
parameterize the helicity angles by the angular distribution:
\begin{equation}
W(\theta_{V_1}, \theta_{V_2}) \propto (1/4)(1- {\cal F}_L) \sin^2 \theta_{V_1} \sin^2 \theta_{V_2} + {\cal F}_L \cos^2 \theta_{V_1} \cos^2 \theta_{V_2}
\end{equation}
\noindent
convolved with a resolution function for each angle. Thus for signal, the PDF is factorized as ${\cal P}_j (m_{V1}^i, m_{V2}^i, \theta_1^i, \theta_2^i) =
 {\cal P}_j (m^i_{V_1}) \times     {\cal P}_j (m^i_{V_2}) \times  W(\theta_{V_1}, \theta_{V_2})$. For the angles  $\psi_{1,2}$, the PDFs are
$\sin^2 \psi_{1,2}$ distributions. For combinatorial background, we use an Argus function~\cite{Argus} for \mes,
a second-order polynomial for \DE\ and a two-piece normal distribution for ${\cal F}$. The masses and
helicity angles of the two vector mesons are considered to be independent and thus are factorized.
We use a third-order polynomial for the $m_V$ distributions. The PDFs of the helicity angles are
third-order polynomials for combinatorial background where the parameterization is obtained
from on-peak sideband data ($m_{ES} < 5.27$~\gevc2). Similarly, we use a third-order polynomial
in $\cos \psi$   for the $\omega$ decay angle. We also consider $B \bar B$ peaking background determined from
simulation. We parameterize \mes, \DE\ and ${\cal F}$ with similar functions as those for the signal.
The PDFs for $m_V$ and $\psi$  angle are similar to those of the combinatorial background. For
the helicity angles, we use a fourth-order polynomial. For the signal, \BB\  and \qq\
background components, we determine the PDF parameters from simulation. We study large
control samples of $B \ra D \pi$ decays of similar topology to verify the simulated resolutions in \DE\
and \mes, adjusting the PDFs to account for any differences found.

\begin{figure}[h]
\centering
\includegraphics[width=50mm]{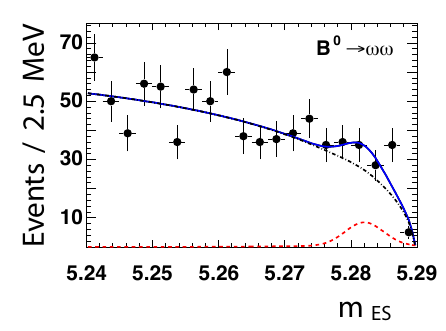}
\includegraphics[width=50mm]{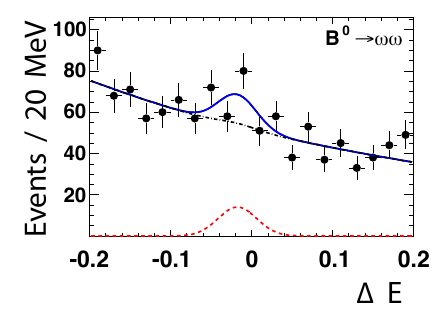}
\includegraphics[width=50mm]{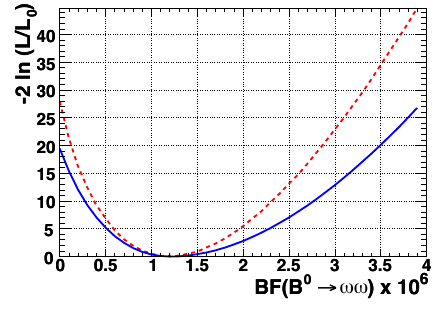}
\includegraphics[width=50mm]{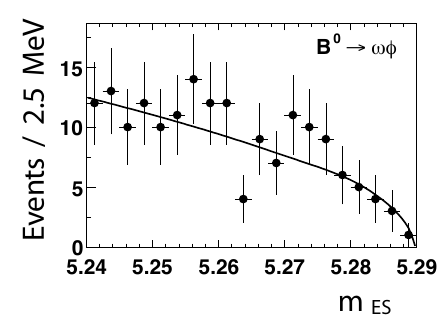}
\includegraphics[width=50mm]{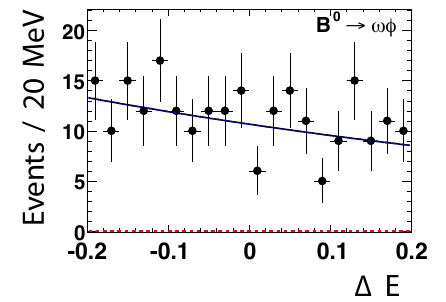}
\includegraphics[width=50mm]{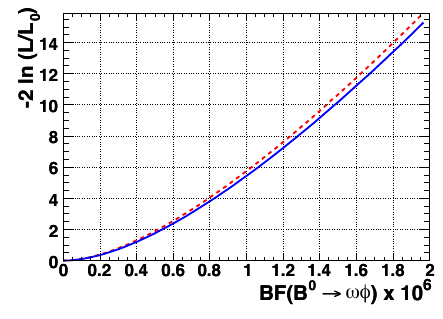}
\vskip -0.3cm
\caption{Fit results for \bomom\ (top row) and \bomfi\ (bottom row) showing  \mes\ (left) and \DE\ (middle)
projections of the  ML fit  for data (points with error bars), the total PDF (solid blue line), background PDF (black dashed line) and signal contribution (red dashed line). The distributions of $-2 \ln ({\cal L}(B)/ {\cal L}_0)$ versus branching fraction (right) for statistical errors only (dashed red curve) and statistical and systematic errors combined (solid blue curve). }
\label{fig:bomom}
\end{figure}

Figures~\ref{fig:bomom} (left, middle) show the \mes\ and \DE\ projections of the multidimensional ML
fit for \bomom\ (top) and \bomfi\ (bottom. Figure~\ref{fig:bomom} (right) shows the corresponding distributions of  
$-2 \ln ({\cal L}(B)/ {\cal L}_0)$ for zero signal normalized to the likelihood at the minimum value
for the two modes. We fix ${\cal F}$ to 0.88 but vary it from 
0.58--1.0 in the systematic error determination. Signal and background yields are extracted from the fit. For \bomom, we observe $N_{\omega \omega}=69^{+16.4}_{-15.2}$ signal events. Including systematic uncertainties,
this yields a $4.4\sigma$ effect. For \bomfi, we see no signal. The fit yields $N_{\omega \phi}= -2.8^{+5.7}_{-4.0}$
signal events. Besides the uncertainty from the \FL\ variation, systematic uncertainties include
contributions from the PDF parameterization, selection efficiency, use of control samples, number
of $B$ mesons and $B$-daughter branching fractions. The largest systematic error comes from yield
bias estimation in the fit yielding order ${\cal O}$(5 events). We measure a branching fraction of ${\cal B}(B \ra \omega \omega) = (1.2\pm 0.3^{+0.3}_{-
0.2}) \times 10^{-6}$, which agrees well with the SM prediction. We do not have enough data to measure \FL\ in this mode. 
For the \bomfi\ mode, we set a Bayesian upper limit of ${\cal B}(B\ra \omega \phi) = 0.7 \times 10^{-6} $ at 90\% CL assuming a  flat prior.

\section{Conclusions}
\label{chap:conclusion}

We have performed the most precise direct \CP\ asymmetry measurement in the semi-inclusive \bsg\ decay.
The measurement error is still suffiiciently large allowing to accommodate new physics contributions in the Wilson coefficient  \Cseven.
We performed the first determination of \imces\ by  measuring the difference in \CP\ asymmetries between charged and neutral $B$ decays. We
measured partial branching fractions and \CP\ asymmetries in \bsll\ that agree well with the SM predictions. We find no evidence for \bpell\ decays
and set branching fraction upper limits at 90\% CL that are a factor of two to three above the SM predictions. We also find no evidence
for lepton number violation in \bhll\  decays and set stringent branching fraction upper limits at 90\% CL. We find first  evidence for \bomom\ decays and measure a branching fraction that is consistent with the SM prediction. However, the data sample is too small to extract ${\cal F}_L$ from an angular analysis. We set an improved branching fraction upper limit at 90\% CL for \bomfi. Significant improvement on
these measurements are expected to come from Belle II and for some decays also from LHCb.

\section{Acknowledgment}
This work has been supported by the Norwegian Research Council. I would like to thank the \babar\ collaboration for the opportunity to give this talk. In particular, I would like to thank Justin Albert, Frank Porter, Fergus Wilson  for useful comments.

\smallskip

\end{document}